\documentclass [11pt]{article}
\usepackage{amsmath}
\usepackage{amssymb}
\usepackage{amscd}
\usepackage{amsthm}
\usepackage{amsopn}
\usepackage{xspace}
\usepackage{verbatim}
\usepackage{amsmath}
\usepackage{amsfonts}
\newtheorem{lemma}{Lemma}[section]
\newtheorem{proposition}{Proposition}[section]
\newtheorem{corollary}{Corollary}[section]
\newtheorem{definition}{Definition}[section]
\newtheorem{acknowledgment*}{Acknowledgment}

\numberwithin{equation}{section}

\newcommand{\up}{\underline{\phi}}
\newcommand{\op}{\overline{\phi}}
\newcommand{\ups}{\underline{\psi}}
\newcommand{\ops}{\overline{\psi}}

\newcommand{\er}{\eqref}

\newcommand{\be}{\begin{equation}}
\newcommand{\ee}{\end{equation}}
\newcommand{\bd}{\begin{displaymath}}
\newcommand{\ed}{\end{displaymath}}

\newcommand{\eps}{\varepsilon}

\newcommand{\R}{\mathbb R}
\renewcommand{\er}{\eqref}

\newcommand{\p}{{\bf H}_p}
\newcommand{\2}{{\bf H}_2}
\renewcommand{\vec}[1]{\boldsymbol{#1}}

\begin{document}
\LARGE   \begin{center} {\bf Optimal Transportation in the
presence of a prescribed pressure field }
\end{center}\normalsize
\begin{center}{G. Wolansky}\footnote{Department of mathematics, Technion, Haifa 32000, Israel}\end{center}
\begin{center}\today\end{center}
\begin{abstract}
The optimal (Monge-Kantorovich) transportation problem is discussed
from several points of view. The Lagrangian formulation extends the
action of the  {\em Lagrangian}  $L(v,x,t)$ from the set of orbits in  $\R^n$  to a set  of measure-valued
orbits. The {\em Eulerian}, dual formulation leads an optimization problem on the set of sub-solutions of the
corresponding Hamilton-Jacobi equation. Finally, the Monge problem  and its
Kantorovich relaxation are obtained by reducing the optimization
problem to the set of measure preserving mappings  and two point
distribution measures subjected to an appropriately defined cost function.
\par
In this paper we concentrate on mechanical Lagrangians
$L=|v|^2/2+P(x,t)$ leading, in general, to a non-homogeneous cost
function. The main results  yield existence of a unique {\em flow} of
homomorphisms which transport the optimal measure valued orbit of
the extended Lagrangian, as well as the existence of an optimal
solution to the dual Euler problem and its relation to the
Monge- and Kantorovich formulations.
\end{abstract}
\section{Introduction}
\subsection{Motivation}
Consider the  Schrodinger equation on a domain
$\Omega\subseteq\R^n$
\begin{equation}\label{schro} i\hbar\frac{\partial\psi}{\partial
t} = \hbar^2\Delta_x\psi + P\psi \ \ ; \ \ (x,
t)\in\Omega\times[0,T] \end{equation} where $P=P(x,t)$ is a given
potential (pressure field). If $\partial\Omega\not=\emptyset$ we
specify a boundary condition under which equation \er{schro} is
well posed and generate a unitary semigroup. We are interested in
the following question: Given a pair of probability measures
$\mu_0,\mu_1$ on $\Omega$ (say, absolutly continuous subjected to
the densities $\rho_0$, $\rho_1$), can one solve the two point
boundary problem, given only  the data $\rho_0(x)=|\psi|^2(x,0)$
and $\rho_1(x)=|\psi|^2(x,T)$?
\par
This problem is a prototype for a variety of wave equations
admitting a Lagrangian formulation. However, in spite of the
underlying Lagrangian stracture associated with \er{schro}, we do
not yet know wether such a two point boundary value problem is
well posed for this equation. \par We turn, therefore, to the {\it
semi-classical approximation} $\hbar\rightarrow 0$, where $\psi$
is written as $\sqrt{\rho} e^{i\phi/\hbar}$. Then the Schrodinger
equation  is reduced to the Hamilton-Jacobi  (HJ) equation
\be\label{hjint} \phi_t + \frac{1}{2}|\nabla_x\phi|^2 =
P\end{equation} coupled with the continuity equation
\be\label{contin} \rho_t + \nabla_x\cdot\left[
\rho\nabla_x\phi\right] =0 \ . \end{equation} The two point
boundary problem for the Schrodinger equation in the semi
classical approximation is, therefore, reduced to solving
\er{hjint} and \er{contin} under the sole data of the initial and
end probability densities $\rho_0=\rho(,0)$ and $\rho_1=\rho(,T)$,
{\it without} any data on the velocity field $\nabla_x\phi$
whatsoever.
\par
 We  attempt to associate a Lagrangian on the state space of
orbits of probability densities $\rho(,t)$ defined for each
$t\in[0,T]$. Since we wish to generalize the standard Lagrangian
formalism defined on classical orbits
$\overline{x}:[0,T]\rightarrow\Omega$, we shall extend this to a
space of "relaxed orbits" of the from $\mu=\mu_{(t)}(dx) dt$,
where $\mu_{(t)}$ is a Borel probability measure on $\Omega$ for
any $t\in [0,T]$. The special class of {\it deterministic} orbits
is embedded in this set as
 $\mu_{(t)}=\delta_{\overline{x}(t)}$, where
$\overline{x}$ is a classical orbit.
\par
The next step is to define a metric on this set of relaxed orbits
which generalizes the action $\int_0^T
|\dot{\overline{x}}(t)|^2dt$ for deterministic orbits.
 It turns out that the above problem is closely related to the
classical Monge-Kantorovich problem of optimal transportation
subjected to  quadratic cost. An historical  background of this is
given below.
\par
\subsection{Historical Background}
The classical problem of optimal mass transportation was suggested
by Monge in the 18'th century [M]:  given a cost function $c(x,y)$
(originally, $c=|x-y|$) and a  pair of Borel probability measures
$\mu_0$, $\mu_1$ on (say) a common probability space $\Omega$,
minimize $$\int c(x, {\bf T}(x)) \mu_0(dx) \eqno{{(\bf M)}}$$
along all Borel mappings ${\bf T}:\Omega\rightarrow\Omega$ which
transport $\mu_0$ into $\mu_1$ (${\bf T}_\# \mu_0=\mu_1$), namely
\be\label{diez}
 \mu_0\left( {\bf T}^{-1} A\right)=\mu_1(A) \
\ \forall \ \text{Borel sets} \ A\subset \Omega
 \ .
\end{equation} The Monge problem was revived in the last century. In
particular, Kantorovich [K] introduced in 1942 a relaxation,
reducing the Monge problem to a linear programming in a cone  of
two-point distributions over $\Omega$ whose marginals are $\mu_0,
\mu_1$ respectively:
%$$ \int_\Omega \Phi(T(x))\mu_0(dx)=\int_\Omega\Phi(x)\mu_1(dx) \ \
%; \ \ \forall\Phi\in C_b(\Omega)  \ , $$
%or, equivalently,
\par
$$ \min_\lambda \int\int c(x,y)\lambda(dx,dy) \ \ \ ; \
\pi_\#^{(0)}\lambda=\mu_0 \ , \ \pi_\#^{(1)}\lambda=\mu_1
\eqno{({\bf K})} $$  Here $\pi^{(i)}, i=0,1$ are the natural
projections of $\Omega\times\Omega$ on its factors. A particular
attention is given to the {\it Wasserstein metrics} \be
\label{wasser} W_p(\mu_0,\mu_1)= \left[\min_\lambda \int\int
|x-y|^p\lambda(dx,dy) \ \ \ ; \ \pi_\#^{(0)}\lambda=\mu_0 \ , \
\pi_\#^{(1)}\lambda=\mu_1 \right]^{1/p} \end{equation} where
$p\geq 1$.

\par
The most striking advantage of the relaxed problem is that a
minimizer always exists by the compactness of the set of
probability measures (assuming $c$ is continuous and $\Omega$ is
compact). It can be shown [Am] that, if $c$ is continuous and
$\mu_0$ contains no atoms, then the minimum of the Kantorovich
problem coincides with the infimum of the Monge problem. The
existence of an optimal Monge mapping is reduced to existence of
such a minimizer of the Kantorovich problem which is supported on
a graph of a Borel map.
\par
Further progress was achieved in the last few decades. In the late
80's and early 90's Brenier [B] studied the Monge problem with a
quadratic cost $c=|x-y|^2$ on $\R^n$ and showed the existence of a
unique, optimal Monge map provided $\mu_0,\mu_1$ posses finite
second moments and  $\mu_0$ is absolutely continuous with respect
to Lebesgue. Moreover, he showed that this map is the gradient of
a convex function $\Phi$ which solves the {\it dual problem}
\be\label{PhiPsi}
 \inf\left\{ \int_{\R^n} \Phi \mu_0(dx) +\int_{\R^n}\Psi d\mu_1(dx)\right\}
  \ ; \ \ \  \Phi,\Psi\in C(\R^n) \ \ ; \ \ \Phi(x)+\Psi(y)\geq x\cdot y \ \ \forall \ x,y\in\R^n \ , \end{equation}
\par
It turned out that {\it any} map ${\bf T}$ which transports
$\mu_0$ to $\mu_1$ must be of the form $\nabla \Phi\circ {\bf S}$
where ${\bf S}$ perseveres $\mu_0$. This results is interpreted as
a Polar factorization for mappings, generalizing the matrix polar
factorization. Another interpretation of $\nabla\Phi$ is as a
monotone vectorized rearrangement in the class of maps
transporting $\mu_0$ to $\mu_1$. A generalization of this result
also holds for more general, strictly convex (and concave) {\it
homogeneous}  cost functions $c(x,y)=h(x-y)$, using special
definitions of convexity ([GM], [C] and references therein).
\subsection{Objectives and main results}
In general, if $\mu_0$ contains an atom, then there is, in
general, no deterministic mapping ${\bf T}$ of any type which maps
$\mu_0$ into $\mu_1$, so there is no sense to compare the
deterministic Monge problem (M) with the probabilistic Kantorovich
problem (K). However, we may still consider the following
alternative formulation in terms of {\it an optimal flow} with
respect to some family of cost functions $c_{t_1,t_2}=J(x,y,
t_1,t_2)$: \vskip .2in\noindent {\bf(F)} : \ {\it Find a relaxed
orbit $\mu=\mu_{(t)}dt$ and a flow of diffeomorphysms ${\bf
T}_{t_1}^{t_2}:\Omega\rightarrow\Omega$  for \\ . \ \ \ \ \ \ \ \
$t_1,t_2\in (0,T)$ such that
\begin{description}
\item{(i)} \ ${\bf T}_{t_1}^{t_2}$ is the optimal Monge mapping
with respect to $c_{t_1,t_2}$  transporting $\mu_{(t_1)}$ to
$\mu_{(t_2)}$  for any $t_1,t_2\in(0,T)$.
\item{(ii)} \ $\lim_{t\rightarrow 0}\mu_{(t)}=\mu_0$ and $\lim_{t\rightarrow
T}\mu_{(t)}=\mu_1$in the weak sense of measures.
\item{(iii)} \ The limits $\lim_{t\rightarrow T} {\bf T}_{\tau}^{t}=: {\bf
T}_{\tau}^T$ exists uniformly and ${\bf T}_{\tau}^T$is a
continuous mappings  for any $\tau\in (0,T)$.
\end{description}}
It is feasible that, once a solution to the flow problem ${\bf F}$
is provided, a $c_{0,T}$ optimal  solution to the Monge problem
{\bf M} with respect to $\mu_0, \mu_1$ exists by ${\bf T}=
\lim_{\tau\rightarrow 0}{\bf T}_\tau^T$ provided  the later limit
exists as a Borel map.
\par
 Our starting point is the
definition of a norm $||\mu||_p$ of a measure-valued orbit as the
minimal  $\mathbb{L}^p_\mu$-norm of the velocity fields ${\vec v}$
which  satisfy the weak form of the continuity  equation
 \be \left\{ \vec{v}=\vec{v}(x,t) \ ; \ \ \int_0^T\int_\Omega \left[ \phi_t +
{\vec v}\cdot\nabla_x\phi\right]\mu_{(t)}(dx)dt =0\ \ \ ; \ \
\forall \phi\in C_0^1(\Omega\times[0.T]) \right\}\ \ .
\label{ceq}\end{equation}
 and \be \label{musubp}
  \|\mu\|_p:=\left[\inf_{\vec{v}}\int_{\Omega\times [0,T]}
  |\vec{v}|^p\mu_{(t)}(dx)dt\right]^{1/p}\end{equation}
  where the infimum is taken over all $\mu-$measurable vectorfield
  $\vec{v}$ satisfying \er{ceq}.
 Denote the set for which
$||\mu||_p<\infty$ as $\p$. This is a normed cone. In
section~\ref{sec1} we shall indicate some of its properties and
prove a compactness embedding of $\p$ (for $p>1$) in a set of
orbits which satisfies Holder continuity in an appropriate
topology. In particular, the end conditions $\mu_0:=\mu_{(0)}$,
$\mu_1:= \mu_{(T)}$ are uniquely defined for $\mu\in \p$ where
$p>1$.
\par
In the rest of the paper we concentrate on the case $p=2$. The
connection between the cost function $c_{t_1,t_2}=J$ posted in
formulation {\bf (F)} above and the pressure $P$ is as follows:
The function $J=J_P$ is the action associated with the Lagrangian
$$ J_P(x,y,t_1,t_2)=\inf_{\overline{x}} \left\{ \int_{t_1}^{t_2}
\left[\frac{|\dot{\overline{x}}(t)|^2}{2} + P(\overline{x}(t),t)
\right]dt  \ \ \ ; \ \overline{x}:[t_1,t_2]\rightarrow\Omega , \ \
\overline{x}(t_1)=x \ , \overline{x}(t_2)=y\right\} \ . $$ The
main result of this paper, formulated in
section~\ref{mainresults}, reveals a connection between the
following approaches:
\par
\begin{description}
\item{{\bf L}: The Lagrangian approach:}
 Minimize a Lagrangian $L_P$ on
the space of orbits $\mu\in\2$: $${\cal L}:=\inf_{\mu} L_P(\mu)  \
\ ; \ \ L_P(\mu):= \frac{1}{2}||\mu||_2^2 + \int_0^T\int_\Omega P
\mu_{(t)}(dx)dt \ \ , \ \ \mu\in\2 \ \ , \ \ \mu_{(0)}=\mu_0 ,
\mu_{(T)}=\mu_1 \ .  $$
\item{{\bf E}: The Eulerian approach:} Maximize on the set of
velocity potentials $\phi$
 \be \label{d2chract}{\cal E} :=
\sup_\phi \left[ \int_\Omega \phi(x,T) \mu_1(dx) - \int_\Omega
\phi(x,0)\mu_0(dx)\right]\end{equation} where the supremum is
taken in the set of all functions  $\phi=\phi(x,t)$ which are
sub-solutions of the Hamilton-Jacobi (HJ) equation \er{hjint} in a
sense to be defined.
\item{{\bf M}: The Monge approach:} \ Minimize on the set of
mappings verifying \er{diez}
$${\cal M}:= \inf_{\bf T}\left\{  \int_\Omega  J_P(x,{\bf T}(x),0,T) \mu_0(dx) \ \ ; \ \ {\bf T}_\#\mu_0=\mu_1\right\}$$
\item{{\bf K}: The Kantorovich approach:} Minimize on the set of 2-point probability measures with prescribed marginal
 $$ {\cal K}:=\min_\lambda \left\{ \int\int J_P(x,y,0,T)\lambda(dx,dy) \
\ \ ; \ \pi_\#^{(0)}\lambda=\mu_0 \ , \
\pi_\#^{(1)}\lambda=\mu_1\right\}
 $$
\end{description}
\vskip .3in  Our first result reveals   the relation between the
above formulation: If $P\in C^1(\Omega\times [0,T])$ then
$$ {\cal L}={\cal E}={\cal K}$$
holds for {\it arbitrary} (probability, Borel) end measures
$\mu_0,\mu_1$. As discussed above, the Monge problem may not have
a solution at all (e.g., if $\mu_0$ contains an atomic measure and
the set of transporting mappings ${\bf T}_\# \mu_0=\mu_1$ is
empty).
\par
The second part of our main result shows the relation between the
flow problem {\bf (F)} and the Lagrangian formulation {\bf L}.
This is the relation between the optimal velocity field $\vec{v}$
realizing \er{musubp} and the induced flow \be\label{flowv}
\frac{d}{dt}{\bf T}_{t_1}^t(x)= \vec{v}\left( {\bf
T}_{t_1}^t,t\right) \ .
\end{equation}
To elaborate, we shall prove
\begin{description}
\item{1)} \
 There exists
 a minimizer $\mu\in\2$ of {\bf L} which satisfies the end
 conditions. This minimizer may be non-unique.
 \item{2)} \ There exists a maximizer $\psi$ of {\bf E}
 which is a {\it Lipschitz} function on $\Omega\times [0,T]$ and
 satisfies the equation
 \be\label{hjwp} \psi_t +
\frac{1}{2}|\nabla_x\psi|^2  = P\end{equation} {\it almost
everywhere}. Again, such a maximizer may be non-unique.
\item{3)} \ The vector field $\vec{v}=\nabla_x\psi$ is defined
{\it everywhere}
 on some relatively closed set $K_0\subset \Omega\times(0,T)$
which contains the support of {\it any} minimal path $\mu$ of {\bf
L} given by (1).
%\end{description}
%\vskip .3in  We further prove
%\begin{description}
\end{description}
Under some additional assumption on $P$ (see Main Theorem in
section~\ref{mainresults}) we also get
\begin{description}
\item{4)} \ The vector field $\vec{v}=\nabla_x\psi$
is locally Lipschitz continuous on $K_0$.
\item{5)} \ The restriction of  $\vec{v}$ to the support of
{\it any} minimal orbit of ${\bf L}$ is uniquely determined.
\item{6)} \ The flow ${\bf T}$ induced by $\vec{v}$  \er{flowv} leaves  $K_0$  invariant.
\item{7)} \ The flow ${\bf T}_{t_1}^{t_2}$ transports $\mu_{(t_1)}$ to
$\mu_{(t_2)}$ for any  minimizer $\mu$ of {\bf L} and any
$t_1,t_2\in (0,T)$. Moreover, it is an optimal Monge transport
with respect to the action $J_P(\cdot,\cdot,t_1,t_2)$.
\item{8)} \ The maps $\lim_{\tau\rightarrow T}
{\bf T}_t^\tau:= {\bf T}_t^T: \Omega\rightarrow\Omega$ and
$\lim_{\tau\rightarrow 0} {\bf T}_\tau^t:={\bf T}_0^t: \Omega
\rightarrow \Omega$ exist and are continuous for any $t\in (0,T)$.
Moreover, $\left[{\bf T}_t^T\right]_\#$ (res. $\left[{\bf
T}_0^t\right]_\#$) is an optimal Monge map with respect to the
action $J_P(,,t,T)$ (res. $J_P(\cdot,\cdot,0,t)$) transporting
$\mu_{(t)}$ to $\mu_1$ (res. $\mu_0$ to $\mu_{(t)}$).
\item{9)} \ If $\lim_{t\rightarrow T} {\bf T}_0^t:= \overline{T}$  exists as a Borel map,
then $\overline{\bf T}$ transports $\mu_0$ to $\mu_1$ and is an
optimal solution of the Monge problem {\bf M}. In this case
$$ {\cal M}={\cal L}={\cal E}={\cal K}$$
\end{description}
\vskip .3in A particular case is the {\it pressureless} flow
 $P\equiv 0$. Here the optimal potential satisfies  \begin{equation} \psi_t +
\frac{1}{2}|\nabla_x\psi|^2  = 0  \label{HJ0}\end{equation} and
the associated action is $$
J_0(x,y,t_1,t_2)=\frac{|x-y|^2}{2(t_2-t_1)} \ ,
$$ reducing the Monge-Kantorovich problem to the Wasserstein
metric $W_2$ for quadratic costs \er{wasser}. The associated flow,
claimed in (6), is given in this case by $$ {\bf
T}_{t_1}^{t_2}(x)=x+ (t_2-t_1)\nabla_x\psi(x, t_1) \ $$ where
$\nabla_x\psi$ is defined and Lipschitz {\it everywhere}. In
particular it follows that, for a quadratic cost, an optimal Monge
map ${\bf T}_\# \mu_0=\mu_1$ exists and is unique provided
$\nabla_x\psi(x,0)$ is $\mu_0$ measurable.\footnote{Since $\psi(,
0)$ is a Lipschitz function, $\nabla_x\psi(x,0)$ is a measurable
function defined a.e, so we recover the existence of an optimal
map if $\mu_0$ is a continuous w.r to Lebesgue measure.}
  In this case, Brenier representation ${\bf
T}={\bf T}_0^T=\nabla_x\Phi$ of the optimal map [see \er{PhiPsi}
and the proceeding discussion] is recovered via
$$\Phi(x)= x^2/2+T\psi(x,0)\ \ .
$$
The connection between the Monge-Kantorovich problem in the
quadratic case  and the flow problem {\bf L} ($P\equiv 0$), as
well as the dual relation {\bf E} together with the
Hamilton-Jacobi equation \er{HJ0} was indicated by several authors
(see [BB], [BBG])\footnote{I wish to thank Prof. D.  Kinderlehrer for turning my attention to these publications.}
as well as in the excellent monograph of Villani [V]. However, to the best of my knowledge, the
existence and
uniqueness result for the flow ${\bf T}_{t_1}^{t_2}$ {\it without
any regularity assumptions} on the end measures $\mu_0,\mu_1$ is new even in
the case $P\equiv 0$. In fact, the existence and uniqueness of the
flow holds even if there is no  optimal Monge map.
\par In section~\ref{main} we shall start to develop the tools
needed for the proof of our main results. Section~\ref{4.1} deals
with a dual formulation for the norm $\|\mu\|_2$ for an orbit of
measure $\mu=\mu_{(t)}dt\in \2$. It follows that
$$
||\mu||_2= \sqrt{\sup \left[ \frac{(\int\int \phi_t
\mu(dxdt))^2}{\int\int |\nabla_x\phi|^2 \mu(dxdt)}\right]}$$ where
the supremum is taken on the set of test functions
$\phi(x,t)=\phi\in C^1_0( \Omega\times[0,T])$.  An equivalent
definition turns out to be \be\label{h2int}
\frac{1}{2}||\mu||^2_2= \sup_{\phi,
P}\left\{-\int_{\Omega_I}P(x,t)\mu(dxdt) -\int_\Omega\phi(x,0)
\mu_0(dx) + \int_\Omega \phi(x,T) \mu_1(dx)\right\}\ \ee where the
infimum above is on the pairs of "velocity potentials" $\phi\in
C^1(\Omega\times [0,T])$ and "pressures" $P=P(x.t)$ which are
related via the Bernulli-type (or Hamilton-Jacobi) equation
\er{hjint}. In case of a {\it prescribed} pressure $P$ (as in this
paper), this identity reveals the relation between the Lagrangian
formulation {\bf L} and the Eulerian one {\bf E}. In
section~\ref{4.2} we imply a dual formulation to a strict convex
perturbation of the Lagrangian ${\bf L}$, leading to  an
approximation of the Euler formulation ${\bf E}$, to be used in
the proof of the main result.
\par
For the proof of the main result we shall also need a series of
auxiliary Lemmas and definitions related to the Hamilton-Jacobi
equation. In subsection~\ref{hjw1} we list these definitions and
Lemmas,
   concerning forward (maximal), backward (minimal) and
  reversible solutions of the Hamilton-Jacobi equation, which are
  essential to the proof of the main results.
 The proofs of the Lemmas  are given in
subsection~\ref{hjw2}. In \ref{hjw3} we utilize these results for
the proofs of our main Theorem.
\par
In the rest of the paper we shall restrict ourselves to the flat
torus $\Omega=\R^n/ \mathbb{Z}^n$. The reason is that we wish to
avoid compactness problems originated from measures on $\R^n$, on
the one hand, and the boundary conditions for the Hamilton-Jacobi
equation required in case of a bounded domain $\Omega\subset
\R^n$. The flat torus is the simplest example in the sense that it
is compact manifold with no boundary, on the one hand, and it
inherits the Euclidean  geometry  from $\R^n$ on the other.  Any
function (or probability measure) on $\Omega$ is understood as a
periodic function (or periodic, normalized per-period measure) on
$\R^n$,  unless otherwise is {\it explicitly specified}. In
particular, a mapping ${\bf T}:\Omega\rightarrow\Omega$   is
understood as a mapping on the covering $\R^n$ which satisfies
${\bf T}(x+\vec{z})= {\bf T}(x)+\vec{z}$ for any $x\in \R^n$ and
any $\vec{z}\in \mathbb{Z}^n$.

\newpage
\begin{center} {\bf List of symbols and definitions}\end{center}
\begin{itemize}
\item $\Omega:= \R^n/ \mathbb{Z}^n$.
\item $I=[0,T] \ \ ; \ \ I_0=(0,T)$
\item $\Omega_I=\Omega\times I$ , \ \ $\Omega_{I_0}=\Omega\times
I_0$.
\item $LIP_l$ is the set of all locally Lipschitz functions in
$\Omega){I_0}$.
\item ${\cal M}$ is the set of all probability Borel measures
supported in $\Omega$.
\item ${\cal M}_I$ is the set of all Borel probability measures
supported on $\Omega_I$ which are decomposable as $\mu\in{\cal
M}_I \Longleftrightarrow \mu=\mu_{(t)} dt$ where $\mu_{(t)}\in
{\cal M}$ a.e. $t\in I$.
\item if $\mu$ is Lebesgue continuous measure, then $\rho_\mu\in \mathbb{L}^1(\Omega_I)$ is the  density of
 $\mu$.
\item $\pi^{(0)}$ (res. $\pi^{(1)}$) is the natural projection of
$\Omega\times\Omega$ on its first (res. second) factor $\Omega$.
\item For any pair $\mu_0, \mu_1\in {\cal M}$, the Wasserstein-p metric is defined by $$W_p(\mu_0,
\mu_1):=\inf_\lambda\int_\Omega\int_\Omega |x-y|^p \lambda(dxdy)$$
where the infimum is on all probability measures on
$\Omega\times\Omega$ such that $\pi^{(0)}_\#\lambda = \mu_0$,
$\pi^{(1)}_\#\lambda=\mu_1$.
\item $\mathbb{E}_\mu(\psi):=\int_0^T\int_\Omega
\psi(x,t)\mu_{(t)}(dx)dt$. Likewise,
$\mathbb{E}_{\mu_{(t)}}(\psi)=\int_\Omega \psi(x,t)\mu_{(t)}(dx)$.
\item A lifting $\nu$ of $\mu\in {\cal M}_I$ is a Borel measure on
$\Omega_I\times \R^n$ such that
$$\mathbb{E}_\nu(\psi):=\int_0^T\int_\Omega\int_{\R^n}
\psi(x,t)\nu(dxdtdv)= \mathbb{E}_\mu(\psi) \  \ ; \
\mathbb{E}_\nu(\psi_t + v\cdot \nabla_x\psi)=0$$ for all $\psi\in
C^1_0(\Omega_I)$.

\end{itemize}
\vskip .3in
\section{A metric space for measure's orbits}\label{sec1}
We start with the following \par\noindent
\begin{definition}\label{liftdef} \ Let $\mu\in {\cal M}_I$. Then
$\mu\in\p(I, {\cal M})$ if there exists a lifting $\nu$ of $\mu$
such that $\mathbb{E}_\nu(|v|^p)<\infty$. We shall also define the
$\p$ norm of $\mu\in \p$ by: $$ ||\mu||_p = \inf_\nu
\left[\mathbb{E}_\nu(|v|^p)\right]^{1/p}$$ where the infimum is
taken over all liftings of $\mu$. \end{definition}
\begin{lemma}\label{lemma1}
 $\p$ is complete
 and locally compact under the weak $C^*$ topology if $p>1$.
 That is, for any bounded sequence $\mu_n$ in $\p$
 we can extract a subsequence which converges in $ C^*(\Omega_I)$ to some $\mu\in
 \p$. In addition:
 $$\lim_{n\rightarrow\infty} ||\mu_n||_p\geq ||\mu||_p \ . $$
\end{lemma}
\begin{proof}
By definition there exists a set of liftings $\nu_n$ corresponding
to $\mu_n$. Moreover, this sequence can be chosen so that $
\mathbb{E}_{\nu_n}(|v|^p)<C$, so $\nu_n$ and $v\nu_n$ are tight on
$\Omega_I\times\R^n$ (since $p>1$ and $\Omega_I$ is compact).
Hence the weak limit $\nu$ of $\nu_n$  is  a lifting of the weak
limit $\mu$ of $\mu_n$, and $\mathbb{E}_{\nu}(|v|^p)<C$, hence
$\mu\in \p$. The same argument also yields the
lower-semi-continuity of $\p$.
\end{proof}

\begin{lemma}\label{lemma2} If $\mu=\mu_{(t)} dt\in \p$, $p>1$ then the map \  $t\rightarrow
\mu_{(t)}$ is a Holder $(p-1)/p$ continuous function from $ I$
into ${\cal M}$ with respect to the weak ($C^*$) topology equipped
with the Wasserstein-1 norm $W_1$: \be\label{W1} W_1(\mu_0,\mu_1)=
\sup_{|\nabla\phi|\leq 1}\int_\Omega\phi(\mu_1(dx)-\mu_0(dx)) \
.\end{equation}\end{lemma}
\par\noindent
\begin{proof}
We know that an optimal lifting $\nu$ exists for $\mu\in\p$. The
measure $\nu$ can be decomposed, by the Theorem of measure's
decomposition [AFP], into $\nu= \mu_{(t)}(dx)  \nu_{x,t}(dv)dt$,
for $\mu$ a.a. $(x,t)$. We may  define now the velocity field
$$\vec{v}(x,t)= \mathbb{E}_{\nu_{x,t}}(v)$$ for $\mu$  a.a.
$(x,t)$. It follows that $\vec{v}\in \mathbb{L}^p_\mu$ and,
moreover, $$||\mu||_p= \left[\int_{\Omega_I}|\vec{v}|^p
\mu(dxdt)\right]^{1/p} \ . $$

 By assumption:
\begin{equation}\int_I \int_\Omega \frac{\partial\phi}{\partial t}
\mu_{(t)}(dx) dt =- \int_I\int_\Omega \vec{v}\cdot\nabla_x\phi
\mu_{(t)}(dx) dt\label{firstest}\end{equation} where
$\phi=\phi(x,t)$ is in $C_0^1(\Omega_I)$. Let $\phi(x,t) =
h(t)\Phi(x)$ with $\Phi\in C^1(\Omega)$ and $h\in C^1_0(I)$. Then
$ f_\Phi(t):= \int_\Omega\Phi(x)\mu_{(t)}(dx)$ satisfies
$$\int_0^T f_\Phi(t) h^{'}(t) dt = -\int_0^T h(t)\int_\Omega
\nabla\Phi\cdot \vec{v} \mu_{(t)}(dx)dt \ . $$ By Holder
inequality $$ \int_0^T h(t)\int_\Omega \nabla\Phi\cdot \vec{v}
\mu_{(t)}(dx)dt\leq |\nabla\Phi|_\infty\left[ \int_0^T h^q(t)
dt\right]^{1/q}||\mu||_p$$ with $q=p/(p-1)$. It follows that
$f_\Phi\in \mathbb{W}^{1,p}(I)$ and, moreover,
$||f_\Phi||_{1,p}\leq C||\Phi||_{1,\infty}$. This implies the
result by Sobolev imbedding together with the dual formulation  of
the $W_1$ norm  \er{W1}.
\end{proof}
\par

 Given $\mu_0$
and $\mu_1\in {\cal M}$, define the set $$\Lambda_p(\mu_0,\mu_1):=
\left\{ \mu= \mu_{(t)} dt\in \p
 \ \ ; \   \mu_{(0)}= \mu_0 \ \ , \ \  \mu_{(T)}= \mu_1 ;
\right\} \ . $$
\begin{corollary}\label{2.1}
The set $\Lambda_p(\mu_0,\mu_1)$ where $p>1$ is closed and locally
compact in  $C(I;C^*(\Omega))$.
\end{corollary}
\par
Similar versions of the Lemma and Proposition below can be found
in [Am]. We also note that Proposition~\ref{ambrosio} in the case
$p=2$ is a special case of our main Theorem (see
section~\ref{mainresults}).
\begin{lemma}\label{reg} {\bf ( Regularization Lemma):} \ \ If $\mu\in\p$ then
there exists a sequence $\mu^\eps\in\p$ of smooth density so that
$\mu=\lim_{\eps\rightarrow 0} \mu^\eps$ holds in $C^*(\Omega_I)$
and, moreover, $$\lim_{\eps\rightarrow 0} ||\mu^\eps||_p=
||\mu||_p \ \ . $$ In addition, for any $t_0, t_1\in I$,
$$\lim_{\eps\rightarrow 0} W_p(\mu^\eps_{t_0}, \mu^\eps_{t_1})=
W_p(\mu_{(t_0)}, \mu_{(t_1)}) \ . $$
\end{lemma}
\par

 We next consider the relation
between $\p$ and the optimal solution of the Kantorovich problem.

\begin{proposition} \label{ambrosio} Assume  $p \geq 1$. Let $\mu_0,
\mu_1\in {\cal M}$. Then $\Lambda_p(\mu_0, \mu_1)\not=\emptyset$.
and
 $$\inf_{\mu\in\Lambda_p(\mu_0, \mu_1)}||\mu||_p= W_p(\mu_0, \mu_1) \ . $$
 \end{proposition}
The proof is similar to the proof of Theorem 4.2 of Ambrosio [Am]
for the metric case ($p=1$) .
\par
We note that Corollary~\ref{2.1} is {\it not} valid in the case
$p=1$. To see it, consider the measure: $$\mu=\sum_j \alpha_j(t)
\delta_{(x-x_j(t)} dt$$ where $x_j=x_j(t)\in C^1(I;\Omega)$ and
$\alpha_j\in C^1_+(I, \R)$ such that $\sum_j\alpha_j(t)=1$
$\forall t\in I$. We can approximate $\mu$ by a sequence of
measures $\mu_m\in \Lambda_1(\mu_0,\mu_1)$ as follows: For each
$m\in \mathbb{N}$ consider the division $t^{(m)}_k=k/m$, $0\leq
k\leq m$ of $I$. Let $\lambda_{m,k}$ be the optimal solution of
 Kantorovich problem  due to $W_1(\mu_{(t^{(m)}_k)},
\mu_{(t^{(m)}_{k+1})})$, and ${\bf T}_{m,k}^{(t)}:= {\bf Id} +
(t-t_{k}^{(m)})\left[{\bf T}_{m,k}-{\bf Id}\right]
/(t^{(m)}_{k+1}-t^{(m)}_k)$. Define $\mu_m$ as follows:
$$\mu_{m,(t_k)}= \mu_{(t_k)} \ \ ; \ \ \mu_{m,(t)} = {\bf
T}^{(t)}_{m,k,\#}\mu_{m,(t_k)} \ \ ; \ \ t^{(m)}_k\leq t \leq
t^{(m)}_{k+1}  \ . $$ Then, by Proposition~\ref{ambrosio},
$\mu_{m}$ are bounded in ${\bf H}_1$ and $\mu_m\rightarrow \mu$.
However, $\mu\not\in {\bf H}_1$ unless $\alpha_j$ are constants in
$t$. To see it, note that the continuity equation takes the form
 $$0= \sum_j\int_I( \alpha_j(t)\phi_t(x_j(t),t) + v_j(t)\cdot \nabla_x\phi(x_j(t),t)dt
 =\sum_j\int_I-\dot{\alpha}_j\phi(x_j(t),t)+
 [v_j(t)-\dot{x}_j]\cdot\nabla_x\phi(x_j(t),t)dt$$
 where $v_j(t)$ are the velocities attributed to $x_j$. It is
 evident that, unless $\dot{\alpha}_j\equiv 0$,  for any possible choice of $v_j$ one can find
 $\phi=\phi(x,t)$ for which the integral on the right does not
 vanish.
\section{Main results}\label{mainresults}
Let the pressure
  $P=P(x,t)\in C^1(\Omega_I)$  and the associated action:
\be\label{actionP}L_P(\mu):=\frac{1}{2}||\mu||^2_2 +
\int_{\Omega_I} P\mu(dxdt) \ \ \  ; \ \ \mu\in \2 \ .
\end{equation}
 Let us recall the
definition of the action $J_P$: \begin{equation}\label{defaction}
J_P(x,y,t_1,t_2)=\inf_{\overline{x}} \left\{ \int_{t_1}^{t_2}
\left[\frac{|\dot{\overline{x}}(t)|^2}{2} + P(\overline{x}(t),t)
\right]dt  \ \ \ ; \ \overline{x}:[t_1,t_2]\rightarrow\Omega , \ \
\overline{x}(t_1)=x \ , \overline{x}(t_2)=y\right\} \ .
\end{equation}
\par\noindent
{\bf Remark:} \ {\it Note that $J_P$ is not a function on
$\Omega=\R^n/ \mathbb{Z}^n$ in each of the variables $x,y$,
separately. However, for each $\vec{q}\in \mathbb{Z}^n$ and each
$x,y\in \R^n$, $t_1,t_2\in I$,
$J_P(x+\vec{q},y+\vec{q},t_1,t_2)=J_P(x,y,t_1,t_2)$.  }\par
\par
\begin{definition} ({\bf L}) (the relaxed Lagrangian):\label{calL}
$${\cal L}(\mu_0,\mu_1):= \inf_{\mu\in \Lambda_2(\mu_0,\mu_1)}L_P(\mu) \ . $$
\end{definition}
\begin{definition}\label{mongedef} ({\bf M}). (the Monge problem):
$$ {\cal M}(\mu_0,\mu_1):= \inf_{{\bf T}_\#\mu_0=\mu_1} \int_\Omega J_P(x,{\bf T}(x),0,T)
\mu_0(dx) \ . $$
\end{definition}
\begin{definition}\label{Kanterdef} ({\bf K}). (the Kantorovich problem):
$$ {\cal K}(\mu_0,\mu_1):= \inf_{\lambda} \int_\Omega J_P(x,y,0,T)\lambda(dxdy) \  $$
among all probability measures on $\Omega\times\Omega$ with the
same $\Omega$ marginals $\mu_0,\mu_1$.
\end{definition}
We now introduce the Hamilton-Jacobi (HJ) equation
\begin{equation} \label{hjeq}\frac{\partial \phi}{\partial t} +
\frac{1}{2}|\nabla_x\phi|^2 = P \ . \end{equation}  Let us denote
the set of {\it classical sub-solutions} of the H-J equation as
$$\Lambda^*_P:= \ \left\{ \phi\in C^1(\Omega_I) \ \ ; \ \ \phi_t +
\frac{1}{2}|\nabla_x\phi|^2 \leq P \ \forall (x,t) \ \ \text{in} \
\Omega_I \right\} \ . $$ For our purpose we need a generalization
of the concept of a classical sub-solution. The concept of {\it
viscosity sub-solution} (see, e.g. [E]) is too restrictive for us.
So, we define a {\it generalized sub-solution} of the H-J equation
as follows: \par

The set of generalized sub solution of the H-J equation is given
by \begin{multline}\label{lambda*} \overline{\Lambda}^*_P:=
\left\{ \phi\in LIP(\Omega_I) \ ; \ \ \forall \overline{x}\in
C^1(I; \Omega), \right.  \\ \left.
 \frac{d}{dt} \phi(\overline{x}(t),t)\leq
\frac{1}{2}\left|\dot{\overline{x}}(t)\right|^2+
P(\overline{x}(t),t) \ {holds \ for \ Lebesgue\ \ a.e} \ t\in
I\right\}\end{multline} \noindent{\bf Remark (i):} \ Note that
$\phi(\overline{x}(t),t)$ is a Lipschitz function on $I$ if $\phi$
is Lipschitz and $\overline{x}\in C^1(I)$. Hence it is
a.e. differentiable (as a function of $t$) on $I$ by Rademacher's Theorem (see, e.g., [E]).\\
\noindent{\bf Remark (ii):} \ It is not difficult to see that any
classical sub-solution is also generalized sub solution, so
$\Lambda^*_P\subset \overline{\Lambda}^*_P$. The concept of
generalized sub-solution is more general than that of a viscosity
sub-solution. The relation between generalized sub-solutions  and
viscosity (and anti-viscosity) sub-solutions is discussed in
section~\ref{hjw1}.
\par
\begin{definition}\label{cale} ({\bf E}): (The Euler formulation):
$${\cal E}(\mu_0,\mu_1):= \sup_{\phi\in\overline{\Lambda}^*(P)}\left\{ \int_\Omega
\phi(x,T) \mu_1(dx)-\int_\Omega\phi(x,0)\mu_0(dx)\right\} \ . $$
\end{definition}
We now state our main result:
%\begin{theorem}
\par\noindent {\bf Main Theorem:} \\ {\it  Assume $P\in C^1(\Omega_I)$.  For any
$\mu_0$, $\mu_1\in {\cal M}$:
\begin{equation} {\cal K}(\mu_0,\mu_1)= {\cal
L}(\mu_0,\mu_1)={\cal E}(\mu_0,\mu_1)
  \ . \label{d3estim}\end{equation}
There exists  minimizers $\mu\in \Lambda_2(\mu_0,\mu_1)$ of ${\bf
L}$ (Definition~\ref{calL}) and a maximizer
 $\psi\in \overline{\Lambda}^*_P$  of ${\bf E}$
 (Definition~\ref{cale})
such that \be \label{hhL}\psi_t + \frac{1}{2}|\nabla_x\psi|^2 = P
\  \ \ ; \ a.e\ \ on\  \Omega_I \ . \end{equation}  Assume, in
addition, there exists $C(t)>0$ on $I_0$ so that
$P(x,t)-C(t)|x|^2$ is a concave function on $\R^n$ for any $t\in
I_0$. Then, for maximizer $\psi$ of ${\cal E}(\mu_0,\mu_1)$, there
exists a closed set $K\subset \Omega_I$ such that
\begin{description}
\item{i)} \ The restriction of $\psi$ to $K_0:= K\cap\Omega_{I_0}$ is
continuously differentiable, the equality \er{hhL} holds \em{for
any} $(x.t)\in K_0$ and $\nabla_x\psi$ is Locally Lipschitz
continuous on $K_0$.
\item{ii)} \ Let $\vec{v}$ be a Lipschitz extension of
$\nabla_x\psi$  to $\Omega_{I_0}$. Let ${\bf T}={\bf
T}_{t_1}^{t_2}$ be the flow  generated by  ${\vec v}$. Then $K_0$
is invariant under this flow.
 \item{iii)} \ A minimizer $\mu\in \Lambda_2(\mu_0,\mu_1)$ of ${\bf
L}$ is not necessarily unique. However,  any such minimizer  is
supported in $K$ and the vectorfield $\vec{v}=\nabla_x\psi$ is
uniquely defined on the support of any such minimizer.
\item{iv)} Any such minimizer is transported by the flow ${\bf T}$, that is
$$ \left[ {\bf T}_{t_1}^{t_2} \right]_\#\mu_{(t_1)}= \mu_{(t_2)}$$
holds for any $t_1,t_2\in (0,T)$. Moreover, if $t_1=0$ (res.
$t_2=T$) then ${\bf T}_0^{t}:=\lim_{\tau\rightarrow 0}{\bf
T}^t_\tau$ (res. ${\bf T}_{t}^T=\lim_{\tau\rightarrow T}{\bf
T}^\tau_t$) are continuous maps transporting $\mu_0$ to
$\mu_{(t)}$ (res. $\mu_{(t)}$ to $\mu_1$).
\item{v)} \  The map ${\bf T}_{t_1}^{t_2}$   are  optimal  with
respect to the cost function $c(x,y)=J_P(x,y, t_1,t_2)$ and the
measures $\mu_{(t_1)}$, $\mu_{(t_2)}$, where either $t_1\in I$,
$t_2\in I_0$ or $t_1\in I_0$, $t_2\in I$.
\item{vi)} \ If $P\equiv 0$ then the optimal solution $\psi$ of ${\bf E}$ (Definition~\ref{cale}) is in $C^{1,1}_{loc}
(\Omega_{I_0})$. In particular, the flow $\bf T$ can be  defined
{\em anywhere} in terms of $\psi$ as
$$ {\bf T}_{t_1}^{t_2}(x)= x+ (t_2-t_1)\nabla_x\psi(x, t_1) \ , \forall t_1<t_2\in I_0 \  \ , \forall x\in\Omega \ .  $$
\end{description}}
%\end{theorem}

\section{Dual representation}  \label{main}
 The key duality argument for
minimizing convex functionals under affine constraints  is
summarized in the following proposition whose proof is given in
the appendix:
\begin{proposition}\label{propdual}
Let ${\bf C}$ a real Banach space and  ${\bf C}^*$ the its dual.
Denote the duality ${\bf C}\doteqdot{\bf C}^*$ relation by
$<c^*,c>\in\R$. Let ${\bf Z}$ a subspace of \ ${\bf C}$ and $h\in
{\bf C}^*$. Let ${\bf Z}^*\subset {\bf C}^*$ given by the
condition $z^*\in {\bf Z}^*$ iff $<z^*-h,z>=0$ for any $z\in {\bf
Z}$. Let ${\cal F}:{\bf C}^*\rightarrow \R\cup\{\infty\}$ a convex
function and
$$ I:= \inf_{c^*\in {\bf Z}^*} {\cal F}(c^*)  \ . $$
 Assume further that  $\overline{A}_0:= \{ c^*\in {\bf C}^* \ ; \ {\cal F}(c^*) \leq
I\}$ is compact (in the $*-$ topology of ${\bf C}^*$).
\\ Then
$$ \sup_{z\in {\bf Z}}\inf_{c^*\in {\bf C}^*} \left[ {\cal F}(c^*)-
<c^*,z> + <h,z>\right]=I \ . $$
 In particular, {\it both} sides equal $\infty$ if \
${\bf Z}^*=\emptyset$.
\end{proposition}

\subsection{Dual representation of $\2$}\label{4.1}
We shall apply Proposition~\ref{propdual} were the space ${\bf C}$
is   all the continuous functions  $q=q(x,t,v)$ on $\Omega_I\times
\R^n$ subjected to: \begin{equation}\label{znorm}
\|q\|:=\sup_{(x,t,v)\in \Omega_I\times
\R^n}\left\{\frac{|q(x,t,v)|}{1+|v|^2} \right\} < \infty \ .
\end{equation} The dual space ${\bf C}^*$ contains all finite Borel
measures $\nu$ on $\Omega_I\times \R^n$ of finite second moments:
$$\int_{\Omega_I\times \R^n}  |\nu|(dxdtdv)< \infty \ \ ; \ \  \int_{\Omega_I\times \R^n} |v|^2 |\nu|(dxdtdv)< \infty \ . $$
Define the subspaces ${\bf Z}$, ${\bf Z}_0$ of ${\bf C}$ as
$$ {\bf Z}_{0}:= \left\{ z=\phi_t + v\cdot \nabla_x\phi \ \ ; \
\ \phi\in C_0^1 (\Omega_I) \ \right\} \subset {\bf Z}:= \left\{
z=\phi_t + v\cdot \nabla_x\phi \ \ ; \ \ \phi\in C^1
(\Omega_{I_0})\cap LIP(\Omega_I) \right\} \ . $$ Given $\mu_0,
\mu_1\in {\cal M}$, define \ $h_{\mu_0,\mu_1}$ as a linear
functional on ${\bf Z}$ as follows:
\begin{equation}\label{hdef}<h_{\mu_0,\mu_1},z>:= \int_\Omega \phi(x,T)
\mu_{1}(dx)-\int_\Omega \phi(x,0) \mu_{0}(dx) \  \ \ \ \mbox{for}
\ z\in {\bf Z} \ , \end{equation}  (in particular,
$<h_{\mu_0,\mu_1},z>=0$ if $z\in {\bf Z}_{0}$).
\begin{lemma}
The functional $h_{\mu_0,\mu_1}$, so defined, is continuous
(bounded) on ${\bf C}$.
\end{lemma}
\begin{proof}
Let $\lambda$ be a probability distribution on
$\Omega\times\Omega$ so that $\pi^{(0)}_ \#\lambda=\mu_0$,
$\pi^{(1)}_ \#\lambda=\mu_1$. Then
\begin{equation}\label{lambdaint}\int_\Omega \phi(x,T) \mu_{1}(dx)-\int_\Omega
\phi(x,0) \mu_{0}(dx)=
\int\int_{\Omega\times\Omega}\left[\phi(y,T)-\phi(x,0)\right]\lambda(dxdy)
\ .
\end{equation}
Now, for $\zeta(s):=\frac{(T-s)x+sy}{T}$ we obtain
\begin{multline}\phi(y,T)-\phi(x,0)=\int_0^T \frac{d}{ds}\phi\left(
\zeta(s),s\right) ds= \int_0^T \left[ \phi_t +
\frac{y-x}{T}\cdot\nabla_x\phi\right]_{\zeta(s),s}ds \\
=\int_0^T z\left(\zeta(s),s,\frac{y-x}{T}\right)ds
\end{multline}
In particular, $$\left|\phi(y,T)-\phi(x,0)\right|\leq
\max_{(x,t)\in \Omega_I}\max_{|v|\leq
Diam(\Omega/T)}|z(x,t,v)|\leq \|z\|\left[ 1 +
\left(\frac{Diam(\Omega)}{T}\right)^2\right]  \ , $$ where we used
the definition on the norm $\|\cdot\|$ on ${\bf C}$ given by
(\ref{znorm}). The proof follows from (\ref{hdef},\ref{lambdaint})
and since $\lambda$ is a probability distribution on
$\Omega\times\Omega$.
\end{proof}
 The corresponding dual spaces are given by
 \be\label{cordu}{\bf Z}^*_0:= \left\{ \nu\in {\bf C}^* ; \
\int_{\Omega_I\times\R^n}z(x,t,v)\nu(dxdtdv) = 0 \ ,  \forall  z
\in {\bf Z}_0\right\}\end{equation} $$ \supset {\bf
Z}^*_{\mu_0,\mu_1}:= \left\{ \nu\in {\bf C}^* ;
\int_{\Omega_I\times\R^n}z(x,t,v)\nu(dxdtdv) = <h_{\mu_0,\mu_1},z>
\ , \forall z\in {\bf Z}\right\}  \ . $$

For any $\mu\in \2$, a convex subset of ${\bf C}^*$ is given by
$$ {\bf C}^*_\mu:= \left\{ \nu\in{\bf C}^* \
; \int_{\Omega_I\times\R^n} \phi(x,t)\nu(dx,dt , dv) =
\int_{\Omega_I}\phi(x,t)\mu(dxdt) \ \ \forall \ \phi\in
C(\Omega_I) \  \right\} \ .
$$
Finally,  $F_\mu:{\bf C}^*\rightarrow \R\cup\{\infty\}$ is defined
by \
$$F_\mu(\nu) = \left\{\begin{array}{cc}
  \frac{1}{2}\int_{\Omega_I\times \R^n} |v|^2\nu(dxdtdv) & \mbox{if} \ \nu\in {\bf C}^*_\mu \\
  \infty & \mbox{if} \ \nu\not\in {\bf C}^*_\mu
\end{array}\right.$$
We obtain
\begin{lemma}\label{l4.1}
The function $F_\mu$ is convex on ${\bf C}^*$ for any $\mu\in\2$.
In addition, if $F_\mu(\nu)<\infty$ and $\nu\in {\bf Z}_0^*$ then
$\nu$ is a lifting of $\mu$. Similarly, if $F_\mu(\nu)<\infty$ and
$\nu\in {\bf Z}_{\mu_0,\mu_1}^*$ then
$\mu\in\Lambda_2(\mu_0,\mu_1)$.
\end{lemma}
\begin{proof}
The proof of Lemma~\ref{l4.1} is almost evident from the
definitions. Let us just prove the last part. Since ${\bf
Z}^*_{\mu_0,\mu_1}\subset{\bf Z}^*_0$ it follows that $\nu$ is a
lifting of  $\mu\in\2$. We only have to show that
$\mu\in\Lambda_2(\mu_0,\mu_1)$. Let  $\phi\in
C^1(\Omega_{I_0})\cap LIP(\Omega_I)$, $\eta=\eta(t)\in C^1_0(I)$
satisfies $0\leq \eta\leq 1$ on $I$ and, for some $\eps>0$,
$\eta(t)=1$ for $\eps\leq t\leq T-\eps$, and $\eta_t\geq 0$ on
$[0,\eps]$, $\eta_t\leq 0$ on $[T-\eps,T]$. Set
$\phi^{(\eps)}=\phi$ on $\Omega\times[\eps,T-\eps]$ and
$\phi^{(\eps)}(x,t)=\phi(x,\eps)$ on $t\in[0,\eps]$ (res.
$\phi^{(\eps)}(x,t)=\phi(x,T-\eps)$ on $t\in[T-\eps,T]$).
 Then
$\eta\phi^{(\eps)}\in C^1_0(\Omega_I)$, so
$$0=\int_{\Omega_I\times \R^n} \left[ (\eta\phi^{(\eps)})_t + \eta
v\cdot\nabla_x\phi^{(\eps)}\right]\nu(dxdtdv)=\int_\eps^{T-\eps}\int_{\Omega\times\R^n}
\left[ \phi_t +  v\cdot\nabla_x\phi\right]\nu_{(t)}(dxdv)dt $$
$$+\int_0^\eps\int_{\Omega}
 \eta_t\phi(x,\eps) \mu_{(t)}(dx)dt+\int_{T-\eps}^T\int_{\Omega}
 \eta_t\phi(x,T-\eps) \mu_{(t)}(dx)dt$$ \be\label{s1}+ \int_0^\eps\int_{\Omega\times\R^n}
\eta v\cdot\nabla_x\phi(x,\eps)
\nu_{(t)}(dxdv)dt+\int_{T-\eps}^T\int_{\Omega\times\R^n} \eta
v\cdot\nabla_x\phi(x,T-\eps) \nu_{(t)}(dxdv)dt \ . \end{equation}
Since $\nu$ is a lifting of some $\mu\in\2$ it follows that
$\nu_{(t)}(dxdv)$ is a probability measure on $\Omega\times\R^n$.
By the Cauchy-Schwartz inequality we estimate the last two
integrals by $2\|\nabla_x\phi\|_\infty
\sqrt{\mathbb{E}_\nu(|v|^2)} \eps^{1/2}$. By Lemma~\ref{lemma2},
$\mu_{(t)}$ is Holder continuous of exponent 1/2 in $t$, with
respect to the $W_1$ topology, so $$\int_0^\eps\int_{\Omega}
 \eta_t\phi(x,\eps) \mu_{(t)}(dx)dt=\int_0^\eps\int_{\Omega}
 \eta_t\phi(x,\eps)  \mu_{(0)}(dx)dt+ O(\eps^{1/2})\|\nabla_x\phi\|_\infty\int_0^\eps|\eta_t|dt$$
\be\label{s2}
 = \int_\Omega\phi(x,\eps)\mu_{(0)}(dx)+ O(\eps^{1/2})\|\nabla_x\phi\|_\infty\int_0^\eps\eta_tdt
 = \int_\Omega\phi(x,\eps)\mu_{(0)}(dx)+O(\eps^{1/2})\|\nabla_x\phi\|_\infty
 \ ,
 \end{equation}
 using $\eta_t\geq 0$ on $[0,\eps]$, hence $\int_0^\eps|\eta_t|=\int_0^\eps\eta_t=1$.
Similarly \be\label{s3}\int_{T-\eps}^T\int_{\Omega}
 (\eta\phi)_t \mu_{(t)}(dx)dt= -\int_\Omega\phi(x,T-\eps)\mu_{(T)}(dx)+O(\eps^{1/2})\|\nabla_x\phi\|_\infty \ .
 \end{equation}
Letting $\eps\rightarrow 0$ we obtain from (\ref{s1},
\ref{s2},\ref{s3}):
$$\int_{\Omega_I\times
\R^n} \left[ \phi_t +  v\cdot\nabla_x\phi\right]\nu(dxdtdv)-
\int_\Omega\left[
\phi(x,T)\mu_{(T)}(dx)-\phi(x,0)\mu_{(0)}(dx)\right]=0 \ .
$$
The above is valid for any $\phi\in C^1(\Omega_{I_0})\cap
LIP(\Omega_I)$. Since $\nu\in {\bf Z}^*_{\mu_0,\mu_1}$ by
assumption, it follows that $\mu_{(0)}=\mu_0$ and
$\mu_{(T)}=\mu_1$, hence $\mu\in\Lambda_2(\mu_0,\mu_1)$.
\end{proof}
\begin{corollary}\label{cor1}
If $\mu\in \2$  then
\begin{equation}\frac{1}{2}||\mu||^2_2=-\inf_{\phi\in
C_0^1}\left\{\int_{\Omega_I}(\phi_t+|\nabla_x\phi|^2/2)\mu(dxdt)
\right\}= \frac{1}{2}\sup_{\phi\in
C_0^1}\frac{\left(\int_{\Omega_I}\phi_t
\mu(dxdt)\right)^2}{\int_{\Omega_I}|\nabla_x\phi|^2 \mu(dxdt)} \ .
 \label{phi0}\end{equation} as well as
 $$ -\inf_{\phi\in
C^1(\Omega_{I_0})\cap LIP(\Omega_I)
}\left\{\int_{\Omega_I}(\phi_t+|\nabla_x\phi|^2/2)\mu(dxdt)
+\int_\Omega\phi(x,0) \mu_0(dx) - \int_\Omega \phi(x,T)
\mu_1(dx)\right\}$$ \be = \left\{\begin{array}{cc}
  \frac{1}{2}||\mu||^2_2 & if \ \mu\in\Lambda_2(\mu_0,\mu_1) \\
  \infty & if \ \mu\not\in\Lambda_2(\mu_0,\mu_1)
\end{array} \label{phi}\right. \ . \end{equation}
\end{corollary}
\begin{proof}
Certainly, $F_\mu$ satisfies all the conditions of
Proposition~\ref{propdual}.  Using Lemma~\ref{l4.1} and
Proposition~\ref{propdual} in the definition of $\|\mu\|_2$
(Definition~\ref{liftdef} for $p=2$) we obtain that
$$ \frac{1}{2}\|\mu\|_2^2=\inf_{\nu\in {\bf Z}^*} F_\mu(\nu)=
\sup_{z\in {\bf Z}}\inf_{\nu\in {\bf C}^*} \left(
F_\mu(\nu)-<\nu,z> \right)$$
$$ = \sup_{\phi\in C^1_0(\Omega_I)}\inf_{\nu\in {\bf C}^*_\mu}\int_{\Omega_I\times\R^n}\left[
\frac{1}{2}|v|^2 - \phi_t - v\cdot\nabla_x\phi\right]\nu(dxdtdv)$$
$$= \sup_{\phi\in C^1_0(\Omega_I)}\inf_{\nu\in {\bf
C}^*_\mu}\int_{\Omega_I\times\R^n}\left[
\frac{1}{2}|v-\nabla_x\phi|^2 - \phi_t -
\frac{1}{2}|\nabla_x\phi|^2\right]\nu(dxdtdv)$$
$$ = \sup_{\phi\in C^1_0(\Omega_I)}\left\{-\int_{\Omega_I}\left[
  \phi_t +
\frac{1}{2}|\nabla_x\phi|^2\right]\mu(dxdt)+
\frac{1}{2}\inf_{\nu\in {\bf C}^*_\mu}\int_{\Omega_I\times
\R^n}|v-\nabla_x\phi|^2\nu(dxdtdv)\right\} \ . $$ So, we set
$\nu=\mu\delta_{v-\nabla_x\phi}$ to annihilate the second integral
and obtain the first equality in (\ref{phi0}).
 For the second equality in
(\ref{phi0}) we observe
$$\inf_{\phi\in
C_0^1}\left\{\int_{\Omega_I}(\phi_t+|\nabla_x\phi|^2/2)\mu(dxdt)
\right\}=\inf_{\phi\in
C_0^1}\inf_{\beta\in\R}\left\{\int_{\Omega_I}(\beta\phi_t+\beta^2|\nabla_x\phi|^2/2)\mu(dxdt)
\right\}$$
$$ = \inf_{\phi\in
C_0^1}\left(-\frac{1}{2}\frac{\left(\int_{\Omega_I}\phi_t
\mu(dxdt)\right)^2}{\int_{\Omega_I}|\nabla_x\phi|^2 \mu(dxdt)}
\right)\ .
$$
Finally, we obtain (\ref{phi}) using the constraint ${\bf
Z}^*_{\mu_0,\mu_1}$ for ${\bf Z}^*$ in Proposition~\ref{propdual}.
\end{proof}
\noindent {\bf Example}: \ Let $\mu=\sum_{i=1}^k \beta_k
\delta_{(x-x_k(t))}$ where $x_k:I\rightarrow \Omega$ satisfies
$\int_0^T|\dot{x}_j|^2dt := |\dot{x}_j|_2< \infty$ and
$\beta_j\geq 0$, $\sum_j\beta_j=1$. Then
$$\int_{\Omega_I}\phi_t\mu(dxdt) = \sum_j
\beta_j\int_0^T\frac{\partial\phi}{\partial t}(x_j(t),t)dt $$ and
$$\int_{\Omega_I}|\nabla_x\phi|^2\mu(dxdt) = \sum_j
\beta_j\int_0^T|\nabla_x\phi|^2(x_j(t),t)dt $$ On the other hand,
$$\int_0^T\frac{\partial\phi}{\partial t}(x_j(t),t)dt = \int_0^T\left[\frac{d\phi}{d t}(x_j(t),t)-
\dot{x}_j(t) \cdot \nabla_x\phi(x_j(t),t)\right] dt$$
$$ =-\int_0^T\dot{x}_j(t) \cdot \nabla_x\phi(x_j(t),t) dt$$
 so, by an application (twice) of the Cauchy-Schwartz inequality,
  $$\frac{\left(\int_{\Omega_I}\phi_t\mu(dxdt)\right)^2}{\int_{\Omega_I}|\nabla_x\phi|^2\mu(dxdt)}
 = \frac{\left(\sum_j\beta_j \int_0^T\nabla_x\phi(x_j(t),t)\cdot \dot{x}_jdt\right)^2 }{\sum_j\beta_j\int_0^T
 |\nabla_x\phi(x_j(t),t)|^2dt}
 \leq \sum_j\beta_j \int_0^T |\dot{x}_j|^2dt \ . $$
 In fact, it can be shown that $||\mu||_2^2$ coincides with the
 above sum, and that there exists a maximizing sequence
 $\phi_n(x,t)$ such that $\nabla_x\phi_n(x_j(t),t)\rightarrow
 \dot{x}_j(t)$ for all $j$ and a.e $t\in I$ (even if some of the
 orbits $x_j$ intersect (!)-see [W]).
\subsection{Dual representation of the Lagrangian} \label{4.2}
We shall now define a strong convex perturbation of the Lagrangian
$L_P$  (Definition~\ref{actionP}). Let also $F:\R\rightarrow
\R^+\cup\{\infty\}$ such that \be\label{Fdef}F(q)=\infty \ \
\text{if} \ \ q<0 \ \ ; \ \ F(0)=0 \ \ \ ; \ \ \ cq^\omega < F(q)<
Cq^\omega \ \text{if} \  q>0\end{equation} where
$1<\omega<1+1/(n+1)$ and $c,C>0$.
%\item{ii)} \ The set ${\bf M}$  of all Borel measures on
%$\Omega_I\times \R^n$ which satisfy $\int_{\Omega_I\times\R^n}
%|v|^2|\nu(dxdtdv)<\infty$.
%\item{iii)}
 The functional  ${\cal I}_\eps^P:{\bf C}^*\rightarrow \R\cup \{\infty\}$ is defined by:
\begin{equation}{\cal I}_\eps^P(\nu):=
  \int_{\Omega_I\times \R^n}\eps F\left(f_\nu\right)dxdtdv+
\frac{1}{2}\int_{\Omega_I\times \R^n}|v|^2 \nu(dxdtdv) +
\int_{\Omega_I\times \R^n}P(x,t) \nu(dxdtdv),
\label{iepsj}\end{equation} if $\nu=f_\nu(x,t,v)dxdtdv$  is
absolutely continuous with respect to Lebesgue measure and the
density $f_\nu$ satisfies $F(f_\nu)\in
\mathbb{L}^1(\Omega_I\times\R^n)$. Otherwise ${\cal
I}_\eps^P(\nu)=\infty$.   Note that, since $F(q)=\infty$ for
$q<0$, it follows that
 ${\cal I}_\eps^P(\nu)=\infty$ if $\nu\in{\bf C}^*$ is not a
 non-negative
 measure. However, ${\cal I}_\eps^P$ can attain a finite value also
 for a measure $\nu$ which is not normalized (i.e not a
 probability measure on $\Omega_I\times\R^n$).
\par
 Given
$\mu_0,\mu_1\in {\cal M}$, define
\be\label{underidef}I^P_{\eps}(\mu_0,\mu_1):=
  \inf_{\nu\in{\bf Z}^*_{\mu_0,\mu_1}}{\cal I}_\eps^P(\nu)  \ .
\end{equation}
 Next, we claim
\begin{lemma}\label{lemma3.1}  \ For any $\eps>0$,  $$I^P_{\eps}(\mu_0,\mu_1)\geq {\cal
L}(\mu_0,\mu_1)$$ where ${\cal L}(\mu_0,\mu_1)$ as in
Definition~\ref{calL}.
\end{lemma}
\begin{proof}
First,  we can restrict ourselves to non-negative measures $\nu\in
{\bf Z}^*_{\mu_0,\mu_1}$, since otherwise $\int F(f_\nu)=\infty$
by (\ref{Fdef}). We only have to show that if $\nu\geq 0$ and
$\nu\in{\bf Z}^*_{\mu_0,\mu_1}$ then $\nu$ is  a lifting of some
$\mu\in\Lambda_2(\mu_0,\mu_1)$.
\par
Using Lemma~\ref{l4.1} it is, therefore, enough to prove that
$\nu_t(dxdv)$ is a probability measure on $\Omega\times\R^n$ for
a.e (Borel) $t\in I$. Setting $\phi(x,t)=\eta(t)\in C^1_0(I)$ we
obtain from (\ref{cordu}) that $$\int_I \left( \int_{\Omega\times
\R^n}\nu_t(dxdv)\right)\frac{d \eta}{dt} dt=0$$ for any such
$\eta$. This implies that $\int_{\Omega\times \R^n}\nu_t(dxdv)$ is
constant for a.e. $t\in I$. Since $\nu\geq 0$  it implies that
$\nu_t$ is a constant multiple of some probability measure on
$\Omega\times\R^n$ for a.e. $t\in I$. This constant equals one
since the $\Omega$ marginal of $\nu_t$ is $C^*$ continuous on $I$
by Lemma~\ref{lemma2} and is a probability measure at $t=0$
$(\mu_0$) and $t=T$ ($\mu_1$).
\end{proof}
We now proceed to a dual formulation of the constraint
minimization of ${\cal I}_\eps^p$. Certainly ${\cal I}_\eps^P$
satisfies the assumption on ${\cal F}$ introduced in
Proposition~\ref{propdual}. In fact, it follows that the set $\{
\nu\in {\bf C}^* \ ; \ \ {\cal I}_\eps^P(\nu) < C\}$ is bounded
(and hence $*-$compact) for any real $C$. Then
Proposition~\ref{propdual} and (\ref{underidef}) yield
$$I_\eps^P(\mu_0,\mu_1)= \sup_{z\in {\bf Z}} \inf_{\nu\in{\bf
C}^*}\left[ {\cal I}_\eps^P(\nu) -<\nu,
z>+<h_{\mu_0,\mu_1},z>\right]$$ $$ = \sup_{\phi\in
C^1(\Omega_I)}\inf_{f}\int_{\Omega_I\times R^n}\left[ \eps
F(f)-f\left( \phi_t + v\cdot\nabla_x\phi-
\frac{1}{2}|v|^2-P\right)\right]dxdtdv+ $$
$$\int_{\Omega}\phi(x,T)\mu_1(dx)-\phi(x,0)\mu_0(dx) \ ,$$ where
$\inf_f$ stands for the infimum on all measurable functions on
$\Omega_I\times\R^n$.  Let $$H_\eps(f,\phi):= \int_{\Omega_I\times
R^n} \left[ \eps F(f) -\left(\phi_t + v\cdot\nabla_x\phi-
\frac{1}{2}|v|^2-P\right)f \right]dxdtdv \ . $$ $$ =
\int_{\Omega_I\times R^n}dxdtdv \left[\eps F(f)-\left(\phi_t
+\frac{1}{2}|\nabla_x\phi|^2 -
\frac{1}{2}|v-\nabla_x\phi|^2-P\right)f\right]$$ Let $F^*$ be the
Legendre transform of $F$: $$F^*(\lambda)=\sup_s \left[ s\lambda
-F(s)\right] \ . $$ By our assumption we know that $F^*$ is also
convex and non-negative on $\R$. It satisfies $F^*(\lambda)=0$ for
$\lambda\leq 0$.
 Now,
 $$
\inf_{f} H_\eps(f,\phi)= -\eps \int_{\Omega_I\times \R^n}
F^*\left(\frac{\phi_t +
|\nabla_x\phi|^2/2-|v-\nabla_x\phi|^2/2-P}{\eps}\right)dxdtdv$$ $$
= -\eps^{1+n/2}\int_{\Omega_I\times \R^n} F^*\left(\frac{\phi_t +
|\nabla_x\phi|^2/2-P}{\eps}-\frac{|v|^2}{2}\right)dxdtdv$$

  Let \begin{equation}\label{Gdef}G(s):=
\int_{\R^n}F^*(s-|v|^2/2) dv   \end{equation} and
\be\label{Psidef} \Psi_\eps(\phi):= -\eps^{1+n/2}\int_{\Omega_I}
G\left(\frac{\phi_t + |\nabla_x\phi|^2/2-P}{\eps}\right)dxdt +
\int_{\Omega}\phi(x,T)\mu_1(dx)-\phi(x,0)\mu_0(dx)\end{equation}

We have proved:
\begin{lemma}\label{lemma3.4}
For $\eps>0$ and $\mu_0,\mu_1\in {\cal M}$, $$
I^P_\eps(\mu_0,\mu_1)=\sup_{\phi\in C^1(\Omega_{I_0})\cap
LIP(\Omega_I) }\Psi_\eps(\phi) \ .  $$
 \end{lemma}

We shall also need the following result, whose proof is direct and
omitted:
 \begin{lemma}\label{lemma3.3}
If  $F$ satisfies (\ref{Fdef}) then, for  some constant $c>0$, the
function $G$ defined in (\ref{Gdef}) satisfies $c
q^{\omega/\omega-1} < G(q) < c^{-1} q^{\omega/\omega-1}$. Thus,
the first integral of (\ref{Psidef}) is estimated by $$
-\eps^{1+n/2}\int_{\Omega_I} G\left(\frac{\phi_t +
|\nabla_x\phi|^2/2-P}{\eps}\right)dxdt=
-O(\eps^{-\alpha})\int_{\Omega_I} \left|\phi_t +
\frac{|\nabla_x\phi|^2}{2}-P\right|^s dxdt$$ where
$\alpha=1/(\omega-1) -n/2>0$ and $ s= \omega/(\omega-1) > 1+n$
(c.f. (\ref{Fdef})).
\end{lemma}
We also need:
\begin{lemma} \label{lemma3.2} Let $\mu_0,
\mu_1\in {\cal M}$. Then there exists a connecting orbit
$\mu\in\Lambda_2(\mu_0,\mu_1)$ of finite $\2$ norm  and a lifting
$\nu$  such that both $\mu$  and $\nu$ has  densities in
$\mathbb{L}^p(\Omega_I)$ (res. $\mathbb{L}^p(\Omega_I\times
\R^n)$), where $1\leq p<1+1/n$.
\end{lemma}
In particular. it follows that for such $\nu$ as guaranteed in
Lemma~\ref{lemma3.2}, each of the integrals in (\ref{iepsj}) is
finite. Hence, there exists $C>0$ (independent of $\eps$) and
$\nu\in {\bf Z}^*_{\mu_0,\mu_1}$ such that  ${\cal I}^P_\eps(\nu)
< C $ for {\it any} $\eps >0$. In particular,
$I^P_\eps(\mu_0,\mu_1)<C$ for any such $\eps$ by
(\ref{underidef}). It follows from this, Lemma~\ref{lemma3.1} and
Lemma~\ref{lemma3.4} that
\begin{corollary}  \label{cor4.2} For any $\mu_0,\mu_1\in {\cal M}$ there exists $C>0$ independent of $\eps$ where
 $$C> \sup_{\phi\in
C^1(\Omega_{I_0})\cap LIP(\Omega_I)}\Psi_\eps(\phi) \geq {\cal
L}(\mu_0,\mu_1) \ .
$$
\end{corollary}

  Lemma~\ref{lemma3.2} is a direct result from
Lemma~\ref{lemma3.25} below. For its presentation we define the
space  $\p([t_0,t_1])$ by restricting $\p=\p(I)$ to orbits defined
for a time  interval $[t_0,t_1]$. The norm of
$\mu\in\p([t_0,t_1])$ is denoted by $||\mu||_{2, [t_0,t_1]}$.
Lemma~\ref{lemma3.25} is also used in the proof of
Lemma~\ref{holderd}.
\begin{lemma}\label{lemma3.25}
For any $t_1>t_2\in I$, any path
$\overline{x}=\overline{x}(t):[t_0,t_1]\rightarrow\Omega$ and any
$\alpha >0$
 there exists an orbit
 $\mu_{(t)}dt\in \p ([t_0,t_1])$
with $\mu_{(t_0)}=\delta_{x_0}$, $\mu_{(t_1)}=\delta_{x_1}$ where
$x_i=\overline{x}(t_i)$, $i=0,1$,  such that
$$ ||\mu||_{2,[t_0,t_1]}^2\leq \int_{t_0}^{t_1}\left|\dot{\overline{x}}\right|^2 dt
+C |t_1-t_0|\alpha^{-2}$$ and $\mu_{(t)}(dx)=\rho(x,t)dx$ where
$\rho\in \mathbb{L}^p(\Omega\times[t_0,t_1])$ for any
$p\in[1,1+1/n)$. Moreover
$$ |\rho|_p\leq C(p)  \left[(t_1-t_0)\left(\frac{\alpha}{t_1-t_0}\right)^{n(p-1)}\right]^{1/p}$$
and $$ supp(\rho)\subset \left\{ (x,t)\in\Omega\times[t_0,t_1] \ ;
\ |\overline{x}(t)-x|\leq C\frac{t_1-t_0}{\alpha} \ \ \forall \
t\in[t_0,t_1]\right\} \ . $$ In particular, the choice
$\overline{x}(t)=x_0+\frac{t-t_0}{t_1-t_0} (x_1-x_0)$ yields
$$ ||\mu||_{2,[t_0,t_1]}^2\leq \frac{|x_1-x_0|^2}{t_1-t_0}
+C |t_1-t_0|\alpha^{-2} \ . $$
\end{lemma}
The proof of Lemma~\ref{lemma3.25} is given in the Appendix.
\section{Proof of main results}\label{hjw}
\subsection{On the Hamilton-Jacobi Equation}\label{hjw1}
In this section we introduce some fundamental results  for the
$HJ$ equation
\begin{equation}\label{hjp}\phi_t+\frac{1}{2}|\nabla_x\phi|^2=P \ \ \ (x,t)\in\Omega_I\end{equation}
where $P\in C^1(\Omega_I)$. The book of L.Evans [E] contains a
detailed exposition on the Hamilton-Jacobi equation. However, the
discussion in [E] is restricted to generalized solutions of
viscosity type and for  time independent Hamiltonians, which
excludes the application of backward solutions and  time dependent
pressure $P=P(x,t)$ . The results in this section are all needed
 for the proof of the Main
Theorem in section~\ref{mainresults} \par

 We list below some properties of the action $J_P$ \er{defaction}:
\begin{lemma}\label{jprop} For $P\in C^1(\Omega_I)$, the
action $J_P$ is satisfies the following:
\begin{description}
\item{(a)}  For $\tau_1< \tau_2\in [0,T]$ and $x_1,x_2\in\R^n$, the value of the action $J_P(x_1,x_2,\tau_1,\tau_2)$
 is realized
along a  (possibly not unique)  orbit $\overline{x}$ which
satisfies the equation \be\label{dotdot}
\ddot{\overline{x}}=\nabla_x P(\overline{x}(t),t) \ \ ; \ \
t\in[\tau_1,\tau_2] \ .
\end{equation}
\item{(b)} Assume further that
there exists $C(t)>0$ so that $P(x,t)-C(t)|x|^2$ is a concave
function on $\R^n$ for any $t\in I_0$. Let $\overline{x}$ be an
optimizer orbit connecting  $x_1,\tau_1$ to $x_2, \tau_2$. For any
$y\in\R^n$ and $t\in (0,T)$ $$ J_P(x_1,x_2,\tau_1,\tau_2) -
J_P(x_1,y,\tau_1,t)\geq $$
\be\label{ipp}\dot{\overline{x}}(\tau_2)\cdot (x_2-y)+
\left[P(x_2,\tau_2)-\frac{1}{2}|\dot{\overline{x}}|(\tau_2)\right](\tau_2-t)
- O(|x_2-y|^2) -o(t-\tau_2)\ , \end{equation}  $$
J_P(y,x_2,t,\tau_2) - J_P(x_1,x_2,\tau_1,\tau_2)\leq $$
\be\label{iipp}\dot{\overline{x}}(\tau_1)\cdot (x_1-y) +\left[
P(x_1,\tau_1)-\frac{1}{2}|\dot{\overline{x}}|(\tau_1)\right](\tau_1-t)+
O(|x_1-y|^2)+ o(t-\tau_1) \ . \end{equation}
\item{(c)} \ For any $x_1,y,x_2\in\R^n$,
$t_1<\tau<t_2$
  \be \label{dd1}
J_P(x_1,y,t_1,\tau) +J_P(y,x_2,\tau,t_2)\geq
J_P(x_1,x_2,t_1,t_2)\end{equation} holds .
\item{(d)} \ For any pair $x_1,x_2\in\R^n$ and a triple $t_1
< \tau < t_2$ there exists $y^*\in\R^n$ (possibly non-unique) for
which the equality holds in \er{dd1}: \be \label{dd1eq}
J_P(x_1,y^*,t_1,\tau) +J_P(y^*,x_2,\tau,t_2)= J_P(x_1,x_2,t_1,t_2)
\ . \end{equation} There exists a (possibly non-unique) optimal
orbit $\overline{x}$ connecting $(x_1, t_1)$ to $(x_2,t_2)$ such
that $\overline{x}(\tau)=y$. However, for {\em any} such optimal
orbit, $\dot{\overline{x}}(\tau)$ is determined {\em uniquely}.
\item{ (e)} \ For any $t>t_1$  $x\in \R^n$ and a.e $y\in \R^n$
\be \frac{\partial}{\partial t} J_P(x,y,t_1,t) +
\frac{1}{2}\left|\nabla_y J_P(x,y,t_1,t)\right|^2 = P(y,t) \ .
\label{hjfrompoint}\end{equation}
\end{description}\end{lemma}
\begin{definition}\label{defreversible}
 $\phi(x,t)$ is
 a {\em forward} solution of \er{hjp} iff, for any $x\in\Omega$ and $t_1>t_0\in I$
 $$\phi(x,t_1)= \inf_{y\in\R^n}
\left[J_P(y,x,t_0,t_1) + \phi(y,t_0)\right]  \ \ . \eqno{({\bf
F})}$$ Likewise, $\phi$ is a {\em backward} solution iff
 $$\phi(x,t_0)= \sup_{y\in\R^n}
\left[-J_P(x,y,t_0,t_1) + \phi(y,t_1)\right] \ .  \eqno{({\bf
B})}$$
\end{definition}
\par\noindent
{\bf Remark}:
 \  {\it It follows, by the remark proceeding
\er{defaction}, that the right sides of (F) (res. (B)) defines a
function which is $\mathbb{Z}^n$ periodic on  $\R^n$, namely
defined on $\Omega$, if $\phi(\cdot, t_0)$ (res. $\phi(\cdot,
t_1)$) is a function on $\Omega$. }\vskip .2in
 For the special case of zero-pressure
Hamilton-Jacobi equation, the action is reduced to
$$ J_0(x_1,x_2, t_1,t_2)= \frac{|x_2-x_1|^2}{2(t_2-t_1)}$$ and
definition~\ref{defreversible} reduces to the (original) Hopf-Lax
formula:
\begin{definition}\label{def52}
A forward solution of the pressureless Hamilton-Jacobi equation
$$ \phi_t + \frac{1}{2}|\nabla_x\phi|^2=0$$
satisfies,  for any $t_1>t_0$ and $x\in\Omega$
 $$\phi(x,t_1)= \inf_{y\in\R^n}
\left[\frac{1}{2}\frac{|x-y|^2}{t_1-t_0} + \phi(y,t_0)\right]  \ \
, \eqno{({\bf F_0})}$$ while a backward solution satisfies
 $$\phi(x,t_0)= \sup_{y\in\R^n}
\left[-\frac{1}{2}\frac{|x-y|^2}{t_1-t_0} + \phi(y,t_1)\right] \ \
. \eqno{({\bf B_0})}$$
\end{definition}
A forward (backward) solution can be constructed from an initial
(end) data at $t=0$ ($t=T$) as follows.
\begin{lemma}\label{claim1}
 For any {\em continuous} initial data $\phi(,0)$ on $\Omega$
and $P\in LIP(\Omega_I)$,
 \be \phi(x,t)= \inf_{y\in\R^n}
\left[J_P(y, x,0,t)  +
\phi(y,0)\right]\label{claim1-1}\end{equation} is a forward
solution and satisfies \er{hjp} a.e.  Moreover, $\phi\in
LIP(\Omega\times (0,T])$ and \be
\frac{|\phi(x,t)-\phi(y,t)|}{|x-y|} \leq
\frac{C}{t}\label{claim1-2}\end{equation} where $C$ is a constant
{\em independent} on $\phi(,0)$.
 Likewise, for any continuous end data $\phi(,T)$
 $$\phi(x,t)= \sup_{y\in\R^n}
\left[-J_P(x, y,0,t)  + \phi(y,1)\right]$$ is a backward solution
and satisfies \er{hjp} a.e., $\phi\in LIP(\Omega\times [0,T))$ and
\be \frac{|\phi(x,t)-\phi(y,t)|}{|x-y|} \leq
\frac{C}{T-t}\label{claim1-3}\end{equation} If, in either cases,
the end data $\phi(,0)$ (res. $\phi(,T)$) is Lipschitz on
$\Omega$, then the corresponding forward (backward) solution is in
$LIP(\Omega_I)$.
 \end{lemma}
 Next, we establish the connection between generalized
 sub-solutions, as defined in \er{lambda*}, and forward/backward
 solutions:
\begin{lemma}\label{viscos}
Both forward and backward solutions are generalized sub-solutions
in the sense of \er{lambda*}. A forward (backward) solution is a
maximal (minimal) generalized sub-solution in the following sense:
If $\psi$ is a generalized sub-solution and $\phi$ is a forward
(backward) solution so that $\psi(x,\tau)=\phi(x,\tau)$ for all
$x\in \Omega$ and some $\tau\in I$, then $\phi(x,t)\geq \psi(x,t)$
($\phi(x,t)\leq \psi(x,t)$) for all $x\in\Omega$ and $t\geq \tau$
($t\leq \tau$)  in $I$.
\end{lemma}
An immediate corollary from Lemma~\ref{viscos} is:
\begin{corollary}\label{claim2}
 Let $\phi$ be a forward solution and $\psi$ is
a backward solution on $\Omega_I$. \begin{description} \item{i)} \
If $\phi(x,T)=\psi(x,T)$ holds $\forall x\in\Omega$  then
$\psi(x,t)\leq \phi(x,t)$ $\forall (x,t)\in \Omega_I$. \item{ii)}
\  Similarly, if  $\phi(x,0)=\psi(x,0)$ holds $\forall x\in\Omega$
then $\psi(x,t)\leq \phi(x,t)$ $\forall (x,t)\in \Omega_I$.
\end{description}
\end{corollary}
 \par
 Next, we wish to address the notion of a {\it reversible}
solution:
\begin{definition}\label{revpair}
A reversible pair  $\{\op, \up\}$ where $\op$ ($\up$) is a forward
(backward) solution on $\Omega_I$ such that $\op(x,0)=\up(x,0)$
and $\op(x,T)=\up(x,T)$ for any $x\in\Omega$. By
Corollary~\ref{claim2}, $\op\geq \up$ on $\Omega_I$. For any such
reversible pair we denote the reversibility set of the pair as the
relatively closed set $K_0(\op,\up)\subset \Omega_{I_0}$ given by
$$K_0(\op,\up):= \{ (x,t)\in\Omega_{I_0} \ ; \  \ \op(x,t)=\up(x,t) \}
\subset\Omega_{I_0} $$ Likewise, $$  K_0^{(t)}(\op,\up)=
K_0(\op,\up)\cap [\Omega\times \{t\} ]\ \ \ \text{for any} \ \
t\in (0,T) \ .
$$ If $\op\equiv \up$ then $\phi:= \op=\up$ is called  a {\em
reversible} solution.
\end{definition}
From Corollary~\ref{claim2} we obtain a way to create reversible
pairs. It turns out that, in the case $P\equiv 0$, this  way
yields reversible solutions:
\begin{lemma}\label{createrev}
Given $\phi_0\in LIP(\Omega)$, let $\phi$ be the forward solution
subjected to $\phi(x,0)=\phi_0(x)$. Let $\underline{\psi}$ be the
backward solution subjected to $\underline{\psi}(x,T)=\phi(x,T)$,
and $\overline{\psi}$ the forward solution subjected to
$$\overline{\psi}(x,0)=\underline{\psi}(x,0) \ . $$ Then
$\{\overline{\psi}, \underline{\psi}\}$ is a reversible pair.
Moreover, if $P\equiv 0$ then $\overline{\psi}=\underline{\psi}$
is a reversible solution.
\end{lemma}
The next Lemmas indicate that reversible pairs (in particular,
reversible solutions) are closely related to {\it classical}
solutions of the Hamilton-Jacobi equation.
\begin{lemma}\label{c1}
If $\phi\in C^1(\Omega_I)$ is a classical solution of \er{hjp}
 then $\phi$ is a
reversible solution.
\end{lemma}
Using Lemma~\ref{jprop} we show that the converse of
Lemma~\ref{c1} also holds, in some sense:
\begin{lemma}\label{cc1}
 If $\{ \op, \up\}$ is a reversible
pair then both  $\overline{\phi}$ and $\underline{\phi}$ are
differentiable on  $K_0:=K_0(\op,\up)$ (cf.,
Definition~\ref{revpair}). Moreover, $\nabla\op:= \{ \nabla_x\op,
\op_t\}=\nabla\up:= \{ \nabla_x\up, \up_t\}$  and the H.J equation
is satisfied on this set. If, in addition, $P$ satisfies the
condition of Lemma~\ref{jprop}-(b) then $\nabla\phi$ is locally
Lipschitz continuous on $K_0$ and $\phi$ satisfies \er{hjp}
pointwise  on this set.
\end{lemma}
\begin{lemma}\label{ccc1} Assume $P$ satisfies the condition of
Lemma~\ref{jprop}-(b). Let $\vec{v}(x,t)$ be a Lipschitz extension
of $\nabla_x\phi$ from $K_0$ to $\Omega_{I_0}$. Then the set $K_0$
is invariant with respect to the (unique) flow generated by the
vectorfield $\vec{v}$.
\end{lemma}
\par
Finally, we introduce the two following results, to be needed in
Section~\ref{hjw3}:
\begin{lemma}\label{holderd}
If $\phi\in C^1(\Omega\times [t_0,t_1])$ satisfying $$
\phi_t+\frac{1}{2}|\nabla_x\phi|^2 = P + \xi \ \ \ ; \ \
(x,t)\in\Omega\times[t_0,t_1]$$ where $P,\xi\in
Lip(\Omega\times[t_0,t_1])$, $s>n+1$ and $\|\xi\|_s$ stands for
the $\mathbb{L}^s(\Omega\times[t_0,t_1])$  norm of $\xi$ then, for
any $x_0$, $x_1$ in $\Omega$, any $t_1>t_0$ and any orbit
$\overline{x}=\overline{x}(t):[t_0,t_1]\rightarrow\Omega$
satisfying $\overline{x}(t_0)=x_0$, $\overline{x}(t_1)=x_1$:
$$\phi(x_1, t_1)-\phi(x_0,t_0) \leq
\frac{1}{2}\int_{t_0}^{t_1}\left|\dot{\overline{x}}\right|^2
+\int_{t_0}^{t_1} P(\overline{x}(t),t)dt +
C_1||\xi||^{2\beta}_s(t_1-t_0)^\lambda  +
C_2||P||_{lip}(t_1-t_0)^\eta||\xi||_s^\beta \ , $$
 where $\beta=
\frac{p}{2p+n(p-1)}$, , $p=s*:= \frac{s-1}{s}$,
$\lambda=\frac{2+n-np}{2p+np-n}$,
$\eta=\frac{4p+(n-1)(p-1)}{2p+n(p-1)}$.\end{lemma} From
Lemma~\ref{holderd} and Definition~\ref{defreversible} we also
obtain
\begin{corollary}\label{claim1.5}
Let $\phi\in C^1(\Omega_I)$ be a  solution and $\psi$ a forward
solution of the respective equations on $\Omega_I$:
$$\phi_t+1/2|\nabla_x\phi|^2=P+\xi \ \ \
; \ \ \ \psi_t+1/2|\nabla_x\psi|^2=P$$ such that
$\psi(x,0)=\phi(x,0)$ on $\Omega$. Then $$\psi(x,T)\geq
\phi(x,T)-\left[C_1||\xi||^{2\beta}_s  +
C_2||P||_{lip}||\xi||_s^\beta\right]$$ where $s,\beta$ as defined
in Lemma~\ref{holderd}.
\end{corollary}
%\end{lemma}
\subsection{Proof of auxiliary results}\label{hjw2}
\begin{proof} (of Lemma~\ref{jprop})
We shall only establish part (b), since all the rest follows by
standard arguments.
\par
To establish this part, let us redefine the action $J_P$ in terms
of {\it parameterized} orbits $x(s)\rightarrow (y(s),\tau(s))$,
$s\in [0,1]$. The optimal orbit is denoted $\overline{x}:=
\{\overline{y},\overline{\tau}\}$. The action takes the form
\be\label{acx1x2} J_P(x_1,x_2,\tau_1,\tau_2)= \inf_{y, \tau}
\int_0^1\left[ \frac{1}{2\tau^{'}(s)}\left|y^{'}(s)\right|^2 +
P(y(s),\tau(s))\tau^{'}(s)\right]ds\end{equation} where $.^{'}$
stands for $s$ derivative while $\{y,\tau\}$ are the set of orbits
satisfying $y(0)=x_1$, $y(1)=x_2$, $\tau(0)=t_1$, $\tau(1)=t_2$
and $\tau^{'}>0$ on $[0,1]$ . An optimal orbit $\{ \overline{y},
\tau\}$ satisfies the Euler-Lagrange equations: \be\label{ELQ}
\frac{d}{ds} \left(\frac{1}{\overline{\tau}^{'}}\frac{d
\overline{y}}{ds}\right)=
\overline{\tau}^{'}P_x(\overline{y},\overline{\tau}) \ \ \ ; \ \ \
\frac{d}{ds}\left[\frac{1}{2}
\left|\frac{\overline{y}^{'}}{\overline{\tau}^{'}}\right|^2 -
P(\overline{y}(s), \overline{\tau}(s))\right] =
-P_\tau(\overline{y},\overline{\tau})\overline{\tau}^{'} \ .
\end{equation}
Now let us replace $x_2$ by $y$ and $\tau_2$ by $t$. Deform the
optimal orbit $\{\overline{y}, \overline{\tau}\}$ into $$
\tilde{y}(s)=\overline{y}(s)+ (y-x_2)s \ \ \ ; \ \ \
\tilde{\tau}(s)=\overline{\tau}(s)+ (t-\tau_2)s$$ Now,
$\{\tilde{y},\tilde{\tau}\}$ is an admissible orbit for the action
with end-points $(x_1, \tau_1)$ and $(y,t)$. In addition, our
assumption on $P$ yields: \begin{multline} P(\tilde{y}(s),
\tilde{\tau}(s))\tilde{\tau}^{'}(s) \leq P(\overline{y}(s),
\overline{\tau}(s))\overline{\tau}^{'}(s) +
\overline{\tau}^{'}(s)P_x(\overline{y}(s),
\overline{\tau}(s))(\tilde{y}-\overline{y})$$ \\ +
\overline{\tau}^{'}(s)P_t(\overline{y}(s),
\overline{\tau}(s))(\tilde{\tau}-\overline{\tau})
+(\tilde{\tau}^{'}-\overline{\tau}^{'})P(\overline{y}(s),
\overline{\tau}(s))+
O^2\left(\left|\tilde{y}-\overline{y}\right|\right)
+o(\tilde{\tau}-\overline{\tau}) \ ,
\end{multline}
as well as
$$ \frac{|\tilde{y}^{'}|^2}{2 \tilde{\tau}^{'}} \leq
\frac{|\overline{y}^{'}|^2}{2 \overline{\tau}^{'}} +
\frac{\overline{y}^{'}(\tilde{y}^{'}-\overline{y}^{'})}{
\overline{\tau}^{'}}-\frac{|\overline{y}^{'}|^2(\tilde{\tau}^{'}-\overline{\tau}^{'})}{2
\left(\overline{\tau}^{'}\right)^2} + O^2(|y-x_2|) +
O^2(|t-\tau_2|) \ ,
$$
 so, by substitution of
$\{\tilde{y},\tilde{\tau}\}$, integration by parts, \er{acx1x2}
and \er{ELQ}  $$ J_P(x_1,y,\tau_1,t) \leq \int_0^1\left[
\frac{1}{2\tilde{\tau}^{'}(s)}\left|\tilde{y}^{'}(s)\right|^2 +
P(\tilde{y}(s),\tilde{\tau}(s))\tilde{\tau}^{'}(s)\right]ds$$ $$ =
J_P(x_1,x_2,\tau_1,\tau_2) +
\frac{1}{\overline{\tau}^{'}(1)}\overline{y}^{'}(1)\cdot(y-x_2) +
\left[ P(\overline{y}(1), \overline{\tau}(1))-
\frac{|\overline{y}^{'}|^2}{2(\overline{\tau}^{'})^2}\right](t-\tau_2)+
O^2\left(|y-x_2|\right)+ o(|t-\tau_2|) \ . $$ Using
$\overline{y}(1)=x_2$, $\overline{\tau}(1)=\tau_2$ and
$\overline{y}^{'}/\overline{\tau}^{'}= \dot{\overline{x}}$ we
obtain \er{ipp}. The second inequality \er{iipp} is obtained
similarly.
\par
\end{proof}
\begin{proof} (of lemma~\ref{claim1}).  \\
Assuming $\phi$ given by \er{claim1-1} and  $t_1>t_0\geq 0$ we
have to prove (F) in Definition~\ref{defreversible}, namely
\be\label{htprove} \phi(y,t_1)= \inf_{w\in\R^n}\left[
J_P(w,y,t_0,t_1) + \phi(w,t_0)\right] \ . \end{equation}
 However,
from \er{claim1-1}
$$\phi(y,t_1)= \inf_{z\in\R^n}\left[  J_P(z,y,0,t_1) +
\phi(z,0)\right] \ \ \mbox{and} \ \ \phi(y,t_0)=
\inf_{z\in\R^n}\left[ J_P(z,y,0,t_0) + \phi(z,0)\right] \ . $$
from \er{dd1}, \er{dd1eq}: $$J_P(z,y,0,t_1)=\inf_{w\in\R^n}\left\{
J_P(z,w,0,t_0) + J_P(w,y,t_0,t_1)\right\}$$ so
$$ \phi(y,t_1)=\inf_{z,w\in\R^n}\left\{
J_P(z,w,0,t_0)+J_P(w,y,t_0,t_1)+ \phi(z,0)\right\}$$
$$= \inf_{w\in\R^n} \left\{ \phi(w,t_0)+ J_P(w,y,t_0,t_1)\right\}
\  $$ and \er{htprove} follows. The proof for the backward
equation is completely analogous.
\par
To prove the second part we proceed, as in the proof of
Lemma~\ref{jprop}, to consider the orbit
$$ \tilde{x}(\tau)=\overline{x}(\tau)+ (y-x)\tau/t$$
where $\overline{x}$ is an optimal orbit connecting $(x,t)$ with
$(\overline{x}(0),0)$, namely
$$\phi(x,t)=\int_0^t\left[\left|\dot{\overline{x}}\right|^2/2 +
P(\overline{x}(\tau),\tau)\right]d\tau+ \phi(\overline{x}(0),0)=
J_P(\overline{x}(0), x,0,t)+  \phi(\overline{x}(0),0) \ .
$$
 Since $\tilde{x}(t)=y$ we obtain, using \er{dotdot} and
 integration by parts
$$\phi(y,t)\leq \int_0^t\left[\left|\dot{\tilde{x}}\right|^2/2 +
P(\tilde{x}(\tau),\tau)\right]d\tau+ \phi(\overline{x}(0),0)=
\phi(x,t)+\dot{\overline{x}}(t)\cdot (y-x)+
\frac{|x-y|^2}{2t}(1+O(t^2)) \ .$$ In particular we obtain that
\be\label{liptt} \frac{\phi(y,t)-\phi(x,t)}{|x-y|}\leq
|\dot{\overline{x}}(t)|+ \frac{|x-y|}{2t}(1+O(t^2))\leq
 \frac{C}{t}\end{equation}
where $C=T\max|\dot{\overline{x}}(t)|
 + Diam(\Omega)\left[
\frac{1}{2}+O(T^2)\right]$. Here $\max|\dot{\overline{x}}|(t)$ is
the maximal possible value over all possible optimal orbits
$\overline{x}$. We now observe that there is a bound
$|\dot{\overline{x}}|<D$ for {\it any} optimal orbit
$\overline{x}$ where  $D$  depends only on the bounds of $P$ and
$P_x$ on $\Omega_I$. This follows since there is a bound on the
variation of $\dot{\overline{x}}$ in terms of $\max(|P_x|)$ via
\er{dotdot}, and a bound on the action $J_P$ itself in terms of
$\max (|P|)$ and
$\max\left(\left|\dot{\overline{x}}\right|\right)$
 by its definition.
\par
The reverse inequality of \er{liptt} and the result  for the
backward solutions follows analogously.
\end{proof}

\begin{proof} (of Lemma~\ref{viscos})
Let $\phi$ be a forward solution. We know that $\phi$ is Lipschitz
by Lemma~\ref{claim1}.   If $\phi$ is not a generalized
sub-solution, then there exists $\eps >0$ and a $C^1$ orbit
$\overline{x}$ so that the set of points
$$ \left\{ t; \ \frac{d}{dt}\phi(\overline{x}(t),t)>
\frac{\dot{|\overline{x}|}^2}{2} +
P(\overline{x}(t),t)+\eps\right\}\subset I$$ is of positive
Lebesgue measure in $I$. Let $\tau_0$ be a density point of this
set. Then, for $t_1 < \tau_0<t_2$ sufficiently close, and
$x_1=\overline{x}(t_1)$ ,$x_2=\overline{x}(t_2)$, we obtain
$$ \phi(x_2,t_2)-\phi(x_1,t_1)=
\int_{t_1}^{t_2} \frac{d}{dt} \phi(\overline{x}(t),t) dt >
\int_{t_1}^{t_2}\left[
\frac{\left|\dot{\overline{x}}(t)\right|^2}{2} +
P(\overline{x}(t),t)\right]dt\geq J_P(x_1,x_2,t_1,t_2)$$ which
contradicts Definition~\ref{defreversible}. Hence, a forward
solution is necessarily a generalized sub-solution.
\par
We now  show that  a forward solution $\phi$ is a maximal
generalized sub-solution. Let $\psi$ be a generalized
sub-solution, $(x,t)\in \Omega_I$. Let $\tau<t$ and
$\overline{x}:[\tau, t]\rightarrow \Omega$ be a $C^1$ orbit
satisfying $\overline{x}(t)=x$. Assume $\phi$ be a forward
solution in $\Omega\times[\tau,t]$ such that
$\phi(x,\tau)=\psi(x,\tau)$ on $\Omega$. Then, by definition
$$ \psi(x,t) = \psi(\overline{x}(\tau),\tau)+ \int_\tau^t
\frac{d}{ds}\psi(\overline{x}(s),s) ds \leq
\psi(\overline{x}(\tau),\tau) +  \int_\tau^t\left[
\frac{1}{2}\left|\dot{\overline{x}}(s)\right|^2 +
P(\overline{x}(s),s)ds\right]$$ \begin{equation}\label{nextline}
\leq \phi(\overline{x}(\tau),\tau) +J_P(\overline{x}(\tau),x,
\tau,t) \ . \end{equation} However, the same inequality  survive
if we take the infimum of the r.h.s of \er{nextline} over all such
orbits $\overline{x}$. By definition, this  infimum is nothing but
the value of the forward solution $\phi(x,t)$. This implies the
maximality of $\phi$.  The statement for a backward solution is
proved analogously.
\end{proof}
\begin{proof} (of Lemma~\ref{createrev}): \\ From
Corollary~\ref{claim2} and by construction  it follows that
$\underline{\psi}\leq \phi$ and  $\overline{\psi}\geq
\underline{\psi}$ on $\Omega_I$, while $\overline{\psi}(x,0)=
\underline{\psi}(x,0)$ for $x\in\Omega$. On the other hand, since
both $\overline{\psi}$ and $\phi$ are forward solutions and
$$\overline{\psi}(x,0)=\underline{\psi}(x,0)\leq \phi_0(x)$$ on $\Omega$, then necessarily
$\overline{\psi}\leq \phi$ on $\Omega_I$.
$$ \overline{\psi}(x,T)\leq \phi(x,T)=\underline{\psi}(x.T)$$
and the equality $\overline{\psi}(x,T)=\underline{\psi}(x,T)$
follows.
\par
Let us turn to the case $P\equiv 0$. Set $\psi_0=\psi(\cdot,0)$
(res. $\psi_1=\psi(\cdot,T)$). From (${\bf F_0}$), (${\bf B_0}$)
in Definition~\ref{def52}:
 $$\psi_1(x)= \inf_{y\in\R^n}
\left[\frac{|x-y|^2}{2T} + \psi_0(y)\right]  \ \ ; \ \
 \psi_0(x)= \sup_{y\in\R^n}
\left[-\frac{|x-y|^2}{2T} + \psi_1(y)\right]   \ . $$ In
particular \be\label{prelegendre} \frac{|x-y|^2}{2T} +
\psi_0(x)-\psi_1(y) \geq 0\end{equation} for any $x,y\in\R^n$. We
may now lift the functions $\psi_0, \psi_1$ from the torus
$\Omega$ to periodic functions in $\R^n$, and set
 $$\Psi_0(x)= T\psi_0(x)+\frac{x^2}{2} \ \ ; \ \ \Psi_1(y)= \frac{y^2}{2}-T\psi_1(y)
 \ , \  \ \ x,y\in \R^n$$
 as functions on $\R^n$.
Then \er{prelegendre} is equivalent   to \be\label{legendre}
\Psi_0(x)+\Psi_1(y) - x\cdot y \geq 0\ \ \ ; \ \forall x,y\in\R^n
\ .  \end{equation} The condition of reversible pair is manifested
in
 $\Psi_0,\Psi_1$ being related by the  Legendre
transforms: $$\Psi_0(x)=\sup_y\left[ x\cdot y-\Psi_1(y)\right] :=
\Psi_1^*(x) \ \ \ ; \ \ \ \Psi_1(x)=\inf_y\left[ x\cdot
y-\Psi_0(y)\right]:= \Psi_0^*(x) \ . $$ In particular, both
$\Psi_0, \Psi_1$ are convex functions on $\R^n$.  Recall that the
{\it super-gradient} $\partial_x\Psi$ of a function $\Psi$ is
given by
$$ y\in \partial_x\Psi \Longleftrightarrow \Psi(z)-\Psi(x)\geq
y\cdot (z-x) \ \ ; \forall z\in \R^n$$ and the equality in
\er{legendre} holds iff $y\in\partial_x\Psi_0$ (equivalently,
$x\in
\partial_y\Psi_1$).
\par
Let now $x,y,z\in \R^n$, $\tau\in (0,T)$. The inequality
\begin{equation}\label{ineqxy}\frac{|x-y|^2}{2T} \leq \frac{|x-z|^2}{2\tau} +
\frac{|z-y|^2}{2(T-\tau)}\end{equation} holds, and turns into an
equality iff $(T-\tau) x + \tau y=Tz$. We obtain
$$ \overline{\psi}(z,\tau)=\inf_{y\in\R^n}
\left[\frac{|z-y|^2}{2\tau} + \psi_0(y)\right] \geq
\underline{\psi}(z,\tau)=\sup_{y\in\R^n}
\left[-\frac{|z-y|^2}{2(T-\tau)} + \psi_1(y)\right] \ , $$ \be
\Longrightarrow  \overline{\psi}(z,\tau)-
\underline{\psi}(z,\tau)= \inf_{x,y\in\R^n}
\left[\frac{|z-x|^2}{2\tau}+\frac{|z-y|^2}{2(T-\tau)}+\psi_0(x)-\psi_1(y)
\right] \ . \label{lastterm}\end{equation} Let now $l(z,\tau):= \{
x,y\in \R^n \ ; \ (T- \tau) x + \tau y = Tz\}$. If we restrict the
infimum  in \er{lastterm} to $x,y\in l(z,\tau)$ then, by the
equality in \er{ineqxy}, it is estimated by
$$ \leq  \inf_{x,y\in l(z,\tau)}
\left[\frac{|y-x|^2}{2T}+\psi_0(x)-\psi_1(y) \right]=T^{-1}
\inf_{x,y\in l(z,\tau)}\left[ \Psi_0(x)+ \Psi_1(y) - x\cdot
y\right]\ .
$$ Now, the last term is zero iff there exists a pair $(x,y)$
where both $(T-\tau) x +\tau y = Tz$ and
 $y\in \partial_x\Psi_0$ hold,
namely:
$$ \frac{Tz-(T-\tau) x}{\tau} \in \partial_x\Psi_0
\Longleftrightarrow \frac{T}{\tau}z:=z_\tau \in
\partial_x\Psi_\tau$$ where $\Psi_\tau(x) :=\Psi_0(x) +
\frac{T-\tau}{2\tau} x^2$. Finally, $z_\tau\in\partial_x\Psi_\tau$
iff $z_\tau$ is in the domain of the Legendre transform
$\Psi^*_\tau$. Since $\Psi_0$ is convex, it follows that
$\Psi_\tau$ is {\it strictly} convex and it's Legendre transform
is defined on $\R^n$.
\end{proof}
\begin{proof} (of Lemma~\ref{c1}): \\
 Let $\overline{x}:[t_0,t_1]\rightarrow\R^n$ be any orbit. Then
$$\int_{t_0}^{t_1}\left[ \frac{|\dot{\overline{x}}|^2}{2}
+P(\overline{x}(t),t) \right]dt= \int_{t_0}^{t_1}\left[
\frac{|\dot{\overline{x}}|^2}{2} +\left(\phi_t(\overline{x},t) +
\frac{1}{2}|\nabla_x\phi(\overline{x},t)|\right) \right]dt$$
$$=\int_{t_0}^{t_1}\left[
\frac{1}{2}|\dot{\overline{x}}-\nabla_x\phi(\overline{x},t)|^2
+\left(\phi_t(\overline{x},t)
+\dot{\overline{x}}\cdot\nabla_x\phi(\overline{x},t)\right)
\right]dt = \int_{t_0}^{t_1}\left[
\frac{1}{2}|\dot{\overline{x}}-\nabla_x\phi(\overline{x},t)|^2
+\frac{d}{dt}\phi(\overline{x}(t),t) \right]dt$$
$$ = \frac{1}{2}\int_{t_0}^{t_1}
|\dot{\overline{x}}-\nabla_x\phi(\overline{x},t)|^2dt + \phi(x_1,
t_1)-\phi(x_0,t_0) \ , $$ where $x_0=\overline{x}(t_0)$,
$x_1=\overline{x}(t_1)$. Hence
$$ \phi(x_1,t_1)=\int_{t_0}^{t_1}\left[
\frac{|\dot{\overline{x}}|^2}{2} +P(\overline{x}(t),t) \right]dt+
\phi(x_0,t_0)-\frac{1}{2}\int_{t_0}^{t_1}
|\dot{\overline{x}}-\nabla_x\phi(\overline{x},t)|^2dt  $$ holds
{\it for any} orbit $\overline{x}(t)$. In particular,
$$ \phi(x_1,t_1)\leq\int_{t_0}^{t_1}\left[
\frac{|\dot{\overline{x}}|^2}{2} +P(\overline{x}(t),t) \right]dt+
\phi(\overline{x}(t_0),t_0) \ . $$   Moreover, if $\overline{x}$
is a solution of
$\dot{\overline{x}}=\nabla_x\phi(\overline{x},t)$,
$\overline{x}(t_1)=x_1$, $t_0\leq t\leq t_1$, then the equality
holds. Hence $\phi$ is a forward solution.  By the same way it
follows that $\phi$ is a backward solution, as well.
\end{proof}
\begin{proof} (of Lemma~\ref{cc1}) \\  Let $\tau\in(0,T)$ and
$x\in\Omega$. Since $\op$ is a forward solution and $\up$ is a
backward one, then
 $$ (i) \ \op(x,\tau)-\phi(y_1,0)\leq J_P(y_1,x,0,\tau)
\ \ \ ; \ \ \ (ii) \ \phi(y_2,T)-\up(x,\tau)\leq J_P(x,y_2,\tau,T)
\ \ \ \forall y_1,y_2\in\R^n \ , $$ Moreover, there exists $y_1^*,
y_2^*\in \R^n$ for which
$$ (i^*) \ \op(x,\tau)-\phi(y^*_1,0)= J_P(y^*_1,x,0,\tau) \ \ \ ;
\ \ \  (ii^*) \ \phi(y^*_2,T)-\up(x,\tau)= J_P(x,y^*_2,\tau,T) \ ,
$$ while, for any $x^*\in\Omega$ and $t\in (0,T)$
$$ (i^{**}) \ \op(x^*,t)-\phi(y^*_1,0)\leq J_P(y^*_1,x^*,0,t)
 \ \ \ \ \ \ \ ; \ \  \ \ \ \ \ \ \ (ii^{**}) \
 \phi(y^*_2,T)-\up(x^*,t)\leq
J_P(x^*,y^*_2,t,T) \ .  $$

From $(i^*)$, $(i^{**})$ and \er{ipp}:
$$\op(x,\tau)-\op(x^*,t)\geq
 J_P(y^*_1,x,0,\tau)- J_P(y^*_1, x^*,0,t)\geq$$ \be\label{(a)}
\dot{\overline{x}}_1(\tau)\cdot(x-x^*) + \left[
P(x,\tau)-\frac{1}{2}|\overline{x}_1(\tau)|^2\right](\tau-t)-
O^2(|x-x^*|)- o(|t-\tau|)\end{equation} where
$\overline{x}_1:[0,\tau]\rightarrow \R^n$ is an optimal orbit
realizing the action $J_P(y_1^*,x,0,\tau)$. Similarly, from
$(i^{**})$
 $(ii^{**})$ and \er{iipp} $$\up(x,\tau)-\up(x^*,t)\leq
J_P(x^*,y_2^*,t,T)-J_P(x, y_2^*,\tau,T)\leq$$ \be\label{(b)}
\dot{\overline{x}}_2(\tau)\cdot(x-x^*)+ \left[
P(x,\tau)-\frac{1}{2}|\overline{x}_2(\tau)|^2\right](\tau-t) +
O^2(|x-x^*|)+ o(|t-\tau|)\end{equation} where
$\overline{x}_2:[\tau,T]\rightarrow \R^n$ is an optimal orbit
realizing the action $J_P(x,y_2^*,\tau,T)$.
\par
Let now $(x,\tau)\in K_0(\op,\up)$. Then $\phi(x,\tau):=
\op(x,\tau)=\up(x,\tau)$ and \er{(a)}, \er{(b)} yield
$$ \dot{\overline{x}}_1(\tau)\cdot(x-x^*)+ \left[
P(x,\tau)-\frac{1}{2}|\overline{x}_1(\tau)|^2\right](\tau-t)-
O^2(|x-x^*|)- o(|\tau-t|) $$
$$\leq \op(x,\tau)-\op(x^*,t)\leq$$
$$
\dot{\overline{x}}_2(\tau)\cdot(x-x^*) + \left[
P(x,\tau)-\frac{1}{2}|\overline{x}_2(\tau)|^2\right](\tau-t)+
O^2(|x-x^*|)+ o(|\tau-t|) \ , $$ as well as
$$ \dot{\overline{x}}_1(\tau)\cdot(x-x^*)
+ \left[
P(x,\tau)-\frac{1}{2}|\overline{x}_1(\tau)|^2\right](\tau-t) -
O^2(|x-x^*|)-o(|t-\tau|) $$ $$\leq \up(x,\tau)-\up(x^*,t)\leq
$$
$$
\dot{\overline{x}}_2(\tau)\cdot(x-x^*)  + \left[
P(x,\tau)-\frac{1}{2}|\overline{x}_2(\tau)|^2\right](\tau-t)+
O^2(|x-x^*|)+ o(|t-\tau|) \ . $$

 Now, in order to show that both
$\nabla_x\op(x,\tau)$ and  $\nabla_x\up(x,\tau)$ exist for
$(x,\tau)\in K_0$ we only have to show that
$\dot{\overline{x}}_1(\tau)=\dot{\overline{x}}_2(\tau)$.
 Adding
(i) and (ii) we obtain:
\begin{equation}\phi(y_2,T)-\phi(y_1,0)\leq
J_P(y_1,x,0,\tau)+J_P(x,y_2,\tau,T)
 \label{y1y2}\end{equation}
for {\it any} $y_1,y_2\in\R^n$. From $(i^{*}),(ii^{*})$ we obtain
that equality holds in \er{y1y2} for $y_1^*,y_2^*$:
\begin{equation}\label{y1y2*}\phi(y^*_2,T)-\phi(y^*_1,0)=
J_P(y^*_1,x,0,\tau)+J_P(x,y^*_2,\tau,T) \ . \end{equation} On the
other hand, by pair-reversibility:
$$\phi(y^*_2,T)-\phi(y^*_1,0)\leq J_P(y_1^*,y_2^*,0,T) \ . $$

Together with \er{y1y2*}: \be\label{above}
J_P(y^*_1,x,0,\tau)+J_P(x,y^*_2,\tau,T) \leq J_P(y_1^*,y_2^*,0,T)
\ . \end{equation} Comparing the above  with \er{dd1} we obtain an
equality in \er{above}.  This implies that
$\dot{\overline{x}}_1(\tau)=\dot{\overline{x}}_2(\tau)$ by
Lemma~\ref{jprop}-d. The existence of
\be\label{eqphi}\nabla_x\phi(x,\tau)=
\dot{\overline{x}}(\tau):=\dot{\overline{x}}_1(\tau)=\dot{\overline{x}}_2(\tau)\end{equation}
and of
$$\phi_t(x,\tau)=P(x,\tau)-\frac{1}{2}|\dot{\overline{x}}|^2(\tau)
$$
 follows
   from \er{(a)}, \er{(b)}.
 In particular  the equality \er{hjp} holds for any $(x,\tau)\in K_0$.
 \par
 We now prove the  Lipschitz continuity of $\nabla_x\phi$ on $K_0$.
By Lemma~\ref{jprop} (b) we obtain that for any $t\in (0,T)$ there
exists $C=C(t)$ such that both $J_P(y,x,0,t) -C(t)x^2$ and
$J_P(x,y,t,T)-C(t)x^2$ are concave functions of $x$ for any
$y\in\R^n$. It follows by Definition~\ref{defreversible} (F) that
 $\overline{q}(x):=\overline{\phi}(x,t)-C(t)x^2$, being the infimum of a set of
 concave functions $J_P(y,x,0,t) -C(t)x^2+ \phi(y,0) \ ; \ y\in\R^n$, is concave
 as well. Likewise, from (B) of the same definition we obtain that $\underline{q}(x):=
 \underline{\phi}(x,t)+ C(t)x^2$, being the
 supremum of a set of convex functions $-J_P(x,y,t,T) + C(t)x^2+ \phi(x,T) \ \ ; \ \ y\in\R^n$,
 is convex.
\par
Let $x, x^*\in K_0$ for some $t\in I_0$. Then both $x, x^*$ are
differentiability points of  the convex function $\underline{q}$,
hence
$$ \left<\nabla_x\underline{q}-\nabla_{x^*}\underline{q},
x-x^*\right> \geq 0\ . $$ Using
$\nabla\underline{\phi}=\nabla\phi$ for $x, x^*$ we obtain
\begin{equation} \left<\nabla_x\phi-\nabla_{x^*}\phi, x-x^*\right> \geq
-C\left|x-x^*\right|^2  \ . \label{ues1}\end{equation} The same
argument applied to the concave function $\overline{q}$ yields
\begin{equation}\label{ues2} \left<\nabla_x\phi-\nabla_{x^*}\phi, x-x^*\right> \leq
C\left|x-x^*\right|^2  \ . \end{equation} We may assume, by
shifting and rotating the coordinate system, that
 $x=(0, 0\ldots 0)$
and $x^*=(\zeta, 0 \ldots 0)$.  Then (\ref{ues1}, \ref{ues2})
yield \be \label{yes} \left| \phi_{,x_1}(0,0\ldots 0,t) - \phi_{,
x_1}(\zeta, 0\ldots 0,t)\right|\leq C |\zeta| \ . \end{equation}
It is left to be shown that there exists a constant $C^*$ such
that \be\label{zarich} \left| \phi_{,x_i}(0,0,\ldots 0,t) -
\phi_{, x_i}(\zeta,0 \ldots 0,t)\right|\leq C^* |\zeta|
\end{equation} holds for any $i>1$. By subtracting an affine
function from $\overline{\phi}$ and $\underline{\phi}$ we may
assume   that (for the fixed value of $t$) \be\label{scale1}
\overline{\phi}(0,t)=\underline{\phi}(0,t)=\overline{q}(0)=\underline{q}(0)=0\end{equation}
and \be\label{scale2}
\nabla_x\overline{\phi}(0,t)=\nabla_x\underline{\phi}(0,t)=\nabla_x\overline{q}(0)=\nabla_x\underline{q}(0)=0
\ , \end{equation} hence (\ref{yes}) is reduced to \be\label{yes1}
\left|\phi_{, x_1}(\zeta,0 \ldots 0,t )\right|< C |\zeta| \ ,
\end{equation} and (\ref{zarich}) takes the form
\be\label{zarich1} \left| \phi_{,x_i}(\zeta,0,\ldots
0,t)\right|\leq C^*|\zeta|  \end{equation} for $i>1$. \par  Since
$\overline{q}$ ($\underline{q}$) is concave (convex), then
(\ref{scale1}, \ref{scale2}) implies that \be\label{osu}
\overline{q}\leq 0 \leq \underline{q}\end{equation} everywhere.

To proceed, we use the concavity of  $\overline{q}$ to obtain
\be\label{soff} \overline{q}(\zeta,y\ldots 0
)-\overline{q}(\zeta,0 \ldots 0)\leq
\overline{q}_{,x_2}(\zeta,0\ldots 0) y =\phi_{,x_2}(\zeta,0\ldots
0,t) y  \ .\end{equation} From (\ref{soff}), (\ref{osu}), the
inequality $\overline{\phi}\geq \underline{\phi}$ and the relation
between $\overline{q}$ and $\overline{\phi}$ we obtain $$
\phi_{,x_2}(\zeta,0\ldots 0,t) y  \geq
\overline{q}(\zeta,y,0\ldots 0)= \overline{\phi}(\zeta,y,0\ldots
0)-C(\zeta^2 + y^2)$$ $$ \geq \underline{\phi}(\zeta,y,0\ldots
0)-C(\zeta^2 + y^2)= \underline{q}(\zeta,y,0\ldots 0)-2C(\zeta^2 +
y^2) \geq -2C(\zeta^2 + y^2) \ ,$$ so $$ \phi_{,x_2}(\zeta,0\ldots
0,t) y + 2C|y|^2 \geq -2C\zeta^2 \  $$  holds for {\it any} $y\in
\R$. Take now the minimum of the lhs above with respect to $y$ to
obtain $$ -\frac{\phi^2_{,x_2}}{8C}\geq -2C\zeta^2\longrightarrow
\left| \phi_{,x_2}(\zeta,0,0\ldots 0,t)\right|\leq 4C|\zeta|
 $$
and (\ref{zarich}) holds with $C^*=4C$ where $i=2$. The general
case  of $i>1$ is evident.
\end{proof}
\begin{proof} (of Lemma~\ref{ccc1}): \ Let $(\overline{\phi},\underline{\phi})$ a reversible pair.
We first note that, for any
$(x,t)\in \Omega_{I_0}$,
$$ 0\leq \overline{\phi}(x,t)-\underline{\phi}(x,t) \leq \phi(y,0)
- \phi(z,T) + J_P(y,x,0,t) + J_P(x,z,t,T) \ . $$ However, from
definition of $J_P$
$$ J_P(y,x,0,t) + J_P(x,z,t,T)\leq
\int_0^T\left[\frac{1}{2}\left|\dot{\overline{x}}(s)\right|^2 +
P(\overline{x}(s),s)\right]ds \ , $$ holds for any $C^1$ orbit
$\overline{x}$ which satisfies $\overline{x}(0)=y$,
$\overline{x}(t)=x$ and $\overline{x}(T)=z$. It follows that
$$ 0\leq \overline{\phi}(x,t)-\underline{\phi}(x,t) \leq
\int_0^T\left[\frac{1}{2}\left|\dot{\overline{x}}(s)\right|^2 +
P(\overline{x}(s),s)\right]ds+
\phi(\overline{x}(0),0)-\phi(\overline{x}(T),T)$$ holds for any
$(x,t)\in\Omega_{I_0}$.
\par
As a result we conclude that, if there exists an orbit
$\overline{x}$ crossing the point $(x,t)\in \Omega_{I_0}$ such
that
\begin{equation}\label{minor}\int_0^T\left[\frac{1}{2}\left|\dot{\overline{x}}(s)\right|^2
+ P(\overline{x}(s),s)\right]ds+
\phi(\overline{x}(0),0)-\phi(\overline{x}(T),T)=0 \ ,
\end{equation} then $\overline{\phi}(x,t)=\underline{\phi}(x,t)$,
i.e.  $(x,t)\in K_0$. Moreover, in that case, the {\it entire}
orbit is contained in $K_0$, namely $\cup_{\tau\in
I_{0}}(\overline{x}(\tau),\tau)\subset K_0$. In addition, it must
be a minimal orbit for the action and, by Lemma~\ref{jprop} and
Lemma~\ref{cc1},
$\nabla_x\overline{\phi}(\overline{x}(t),t)=\nabla_x\underline{\phi}(\overline{x}(t),t)=
\dot{\overline{x}}(t) $.
\par
Next, let $(x,t)\in K_0$ and let
$\overline{x}_1:[0,t]\rightarrow\Omega$
($\overline{x}_2:[t,T]\rightarrow\Omega$)  the minimal orbits for
which $\overline{\phi}(x,t)= \phi(\overline{x}_1(0),0) +
\int_0^t\left[ \frac{1}{2}\left|\dot{\overline{x}}_1(s)\right|^2
+P(\overline{x}_1(s),s)\right]ds$ (res. $\underline{\phi}(x,t)=
\phi(\overline{x}_2(T),T) - \int_t^T\left[
\frac{1}{2}\left|\dot{\overline{x}}_2(s)\right|^2
+P(\overline{x}_2(s),s)\right]ds$) and
$\overline{x}_1(t)=\overline{x}_2(t)=x$. Then $\overline{x}=
\overline{x}_1\cup\overline{x}_2$ satisfies \er{minor}. In
particular, it is contained in $K_0$ and is an integral curve of
the vectorfield $\nabla_x\phi$. This implies the invariance of
$K_0$ under the flow of $\nabla_x\phi$.
\end{proof}

%\vskip .2in\noindent

 \subsection{Proof of the main Theorem}\label{hjw3}
 First, the existence of a minimizer for ${\cal L}(\mu_0,\mu_1)$
 in $\Lambda_2(\mu_0,\mu_1)$ follows immediately by the
 lower-semi-continuity of $\|\mu\|_2$ and the local  compactness
 of $\2$. Next,
we shall prove the chain of inequalities: $$ {\cal
E}(\mu_0,\mu_1)\geq {\cal L}(\mu_0,\mu_1)\geq {\cal
K}(\mu_0,\mu_1) \geq {\cal E}(\mu_0,\mu_1)$$ from left to right,
together with the existence of a maximizer for ${\cal
E}(\mu_0,\mu_1)$ in $\overline{\Lambda}_P^*$.
\begin{itemize} \item{${\cal E}(\mu_0,\mu_1)\geq {\cal
L}(\mu_0,\mu_1)$} \\ From Lemma~\ref{lemma3.3} and
Corollary~\ref{cor4.2}  there exists a sequence $\eps_k\rightarrow
0$ and $\phi_k\in C^1(\Omega_I)$ such that
 \be\label{rr1}
{\cal L}\leq  \Psi_{\eps_k}(\phi_k)<
-O\left(\eps_k^{-\alpha}\right)||\xi_k||_s^s + \int_\Omega
\phi_k(x,T)\mu_1(dx)-\int_\Omega\phi_k(x,0)\mu_0(dx)
< C
\end{equation} where $\alpha >0$, $s>n+1$  and $$\xi_k=\phi_{k,t}+|\nabla_x\phi_k|^2/2-P \
  \ . $$
 Let
now $\Xi:\Omega_I\rightarrow\Omega$ be a flow such that
\begin{description}
\item{i)} \ \  $\sup_{x\in\Omega}\int_0^T\left|\frac{\partial \Xi(x,t)}{\partial
t}\right|^2dt := E <\infty$,
\item{ii)} \ $\Xi(x,0)=x$, $\Xi_\#(\cdot,T)\mu_0=\mu_1$
\end{description}
 By Lemma~\ref{holderd} and (i) we obtain
$$  \phi_k(\Xi(x,T),T)-\phi_k(x,0) \leq \frac{1}{2}E +|P|_\infty +
C_1||\xi_k||^{2\beta}_s + C_2||P||_{lip}||\xi_k||_s^\beta \  . $$
Integrate the above against $\mu_0$ on $\Omega$ and use (ii)  to
obtain \be\label{rr2}
\int_\Omega\phi_k(x,T)\mu_1(dx)-\int_\Omega\phi_k(x,0)\mu_0(dx)
\leq \frac{1}{2}E +|P|_\infty + C_1||\xi_k||^{2\beta}_s +
C_2||P||_{lip}||\xi_k||_s^\beta \ \end{equation} Using $s>n+1$ and
$\beta < 1/2$ (c.f. Lemma~\ref{holderd}) we obtain from  \er{rr1}
and \er{rr2}  that $||\xi_k||_s\rightarrow 0$ as
$\eps_k\rightarrow 0$. In addition
\begin{equation} \liminf_{k\rightarrow\infty}\left[\int_\Omega\phi_k(x,T)
\mu_1(dx)-\int_\Omega\phi_k(x,0) \mu_0(dx)\right]\geq {\cal L} \ \label{eqls} \ . \end{equation}
 Now, we may replace the sequence
$\phi_k$ by a sequence of forward solutions $\overline{\psi}_k$ of
the equation
$$\overline{\psi}_{k,t}+\frac{1}{2}\left|\nabla_x\overline{\psi}_k\right|^2
= P
\
\
\
; \ \ \ \overline{\psi}_k(x,0)=\phi_k(x,0) \ . $$ This is also a maximizing
sequence which, by Corollary~\ref{claim1.5} together with
$||\xi_k||_s\rightarrow 0$, yields
 \begin{equation} \liminf_{k\rightarrow\infty}\left[\int_\Omega\overline{\psi}_k(x,T)
\mu_1(dx)-\int_\Omega\overline{\psi}_k(x,0) \mu_0(dx)\right]\geq
{\cal L} \ \label{eqls1} \ . \end{equation}
  \par From Lemma~\ref{claim1} we also
obtain a uniform estimate on $\overline{\psi}_k$ in
$LIP(\Omega\times (0,T])$. In particular, the sequence
$\overline{\psi}_k(\dot,T)$ is uniformly Lipschitz on $\Omega$.
\par Now, define $\underline{\psi}_k$ to be the {\it backward} solutions of
\er{hhL} subjected to
$\underline{\psi}_k(x,T)=\overline{\psi}_k(x,T)$. From the first
part of Lemma~\ref{claim1}, $\underline{\psi}_k(x,0)\leq
\overline{\psi}_k(x,0)$ on $\Omega$ so \er{eqls1} is satisfied for
$\underline{\psi}_k$ as well. Moreover, by the last part of
Lemma~\ref{claim1} $\underline{\psi}_k$ are uniformly bounded in
the Lipschitz norm on $\Omega_I$.  So, we can extract a
subsequence of $\underline{\psi}_k$ which converges  uniformly on
$LIP(\Omega\times[0,T])$ to a backward solution
$\underline{\psi}$.In particular, {\it both} $\underline{\psi}(,
T)$ and $\underline{\psi}(,0)$ are Lipschitz. Let
$\overline{\psi}$ be the forward solution satisfying
$\overline{\psi}(,0)=\underline{\psi}(,0)$.  By
Corollary~\ref{claim2} and definition~\ref{revpair}, the pair
$(\overline{\psi}, \underline{\psi})$ is a reversible pair and
both functions are  in $\overline{\Lambda}^*_P$ (see the first
part of Lemma~\ref{viscos}). Moreover, the inequality \er{eqls1}
is preserved in the limit process, so
$$\int_\Omega\psi(x,T) \mu_1(dx)-\int_\Omega\psi(x,0)
\mu_0(dx)\geq {\cal L}$$ holds for both $\psi=\overline{\psi}$ and
$\psi=\underline{\psi}$ (recall $\underline{\psi}=\overline{\psi}$
on $\Omega\times\{0\}$ and $\Omega\times\{T\}$).
\item{${\cal L}(\mu_0,\mu_1)\geq {\cal K}(\mu_0,\mu_1)$} \\
Recall that there exists a minimizer of ${\cal L}(\mu_0,\mu_1)$ by
the first part of the Theorem.  Let $\mu$ be such a minimizer. We
now use the regularization Lemma~\ref{reg} to approximate $\mu$ by
smooth densities $\mu_n=\rho_n(x,t)dxdt$. Let $\vec{v}_n$ be the
regularized velocity field. Then
\begin{equation}\lim_{n\rightarrow\infty} \int_{\Omega_I}|\vec{v}_n|^2(x,t)\rho_n(x,t) dxdt = ||\mu||_2^2 \  \ .
\label{xxx}\end{equation} as well as
$$\lim_{n\rightarrow\infty}\int_{\Omega_I}\rho_n(x,t)P(x,t)dxdt=
\int_{\Omega_I} P\mu(dxdt) \ . $$
 Define \be\label{vm}m_n(x,t)=
\rho_n(x,t)\vec{v}_n(x,t) \ . \end{equation}  Then $m_n\in
C^1(\Omega_I)$.
 Define now $$\vec{v}^\eps_n(x,t)=
\frac{m_n(x,t)}{\rho_n(x,t)+\eps}$$ By assumption,
$\vec{v}_n^\eps$ is Lipschitz on $\Omega_I$, $t\in I$. Define
$\rho_n^{(\eps)}(x,t)$ as the solution of \begin{equation}
\frac{\partial \rho_n^{(\eps)}}{\partial t} +
\nabla_x\left[\vec{v}_n^{\eps}\rho_n^{(\eps)}\right] =0 \ \ ; \ \
\rho_n^{(\eps)}(x,0)=\rho_n(x,0) \ . \label{eqrhon}\end{equation}
Since $\vec{v}^{\eps}_n$ is Lipschitz, we may define the flow
associated with it as $\Gamma^t_{(\eps)}:\Omega\rightarrow\Omega$
for $t\in I$, namely $\Gamma^t_{(\eps)}(x) = y_{(x)}(t)$ where
$\dot{y}_{(x)} = \vec{v}_n^{\eps}(y_{(x)}(t),t)$ and
$y_{(x)}(0)=x$.  It follows that
$\Gamma^t_{(\eps),\#}\rho_n(\cdot, 0) dx = \rho_n^{(\eps)}(\cdot,
t) dx$ for all $t\in I$. In particular: $${\cal K}\left(
\rho_n(x,0)dx, \rho^{(\eps)}_n(x,T)dx\right) \leq \int_{\Omega}
\rho_n(x,0)J_P\left(x, \Gamma^T_{(\eps)}(x)0,T\right) dx $$ $$=
\int_{\Omega_I} \rho_n(x,0) \frac{d}{d t} J_P\left(x,
\Gamma^t_\eps(x),0,t\right) dt dx$$ \be = \int_{\Omega_I}
\rho_n(x,0) \left[ \frac{\partial}{\partial t}
J_P(x,\Gamma^{t}_{(\eps)}(x),t) +
\nabla_{y=\Gamma^t_\eps(x)}J_P(x,y,0,t) \cdot
\vec{v}_n^\eps\left(\Gamma^t_\eps(x),t\right)\right]
dxdt\label{532}\end{equation} From \er{hjfrompoint} $$\partial_t
J_P(x,y,0,t) + \nabla_yJ_P(x,y,0,t)\cdot \vec{v}_n^\eps(y,t) =
P(y,t) + \frac{1}{2} \left|\vec{v}_n^\eps\right|^2(y,t) -
\frac{1}{2}\left| \nabla_y J_P(x,y,0,t) -
\vec{v}_n^\eps(y,t)\right|^2 \ . $$ Substitute the above in
\er{532} at $y=\Gamma^t_\eps(x)$ to obtain $${\cal K}\left(
\rho_n(x,0)dx, \rho^{(\eps)}_n(x,T)dx\right) \leq  \int_{\Omega_I}
\rho_n(x,0) \left[ P\left(\Gamma^t_\eps(x),t\right) + \frac{1}{2}
\left|\vec{v}^n_\eps\right|^2\left(\Gamma^t_\eps(x),t\right)\right]dxdt
$$ \be\label{iiijk} = \int_{\Omega_I} \rho_n^{(\eps)}(x,t) \left[
\frac{1}{2} \left|\vec{v}^n_\eps\right|^2(x,t)+ P(x,t)\right] dxdt
\leq \int_{\Omega_I} \rho_n^{(\eps)}(x,t) \left[ \frac{1}{2}
\left|\vec{v}^n\right|^2(x,t)+ P(x,t)\right] dxdt \
\end{equation} where the last inequality follows from \er{vm}. We
next show that
\begin{equation}\lim_{\eps\rightarrow 0}
\int_{\Omega}\left|\rho_n^{(\eps)}(x,t)-\rho_n(x,t)\right|dx
=0\label{limeq}\end{equation} for any $t\in I$. In fact, we note
that $\rho_n+\eps$ solves equation (\ref{eqrhon}), hence
$w_n^{(\eps)}:=\rho_n-\rho_n^{(\eps)} +\eps$ solves this equation
as well. Since $w_n^{(\eps)}(x,0)=\eps >0$  we obtain that
$w_n^{(\eps)}\geq 0$ over $\Omega_I$ and, moreover,
$$\int_{\Omega}\left|\rho_n(x,t)-\rho_n^{(\eps)}(x,t)\right|dx-|\Omega|\eps
\leq  \int_{\Omega}\left|w_n^{(\eps)}(x,t)\right| =
\int_{\Omega}\left|w_n^{(\eps)}(x,0)\right|= |\Omega|\eps$$ for
all $t\in I$. Now we take first the limit $\eps\rightarrow 0$ then
the limit $n\rightarrow\infty$ in \er{iiijk}. The r.h.s of
\er{iiijk} converges to ${\cal L}(\mu_0,\mu_1)$. Now,
$\rho_n(x,0)dx$ and $\rho_n^\eps(x,T)dx$ converges, as
$n\rightarrow\infty$ and $\eps\rightarrow 0$,  weak$-*$ to $\mu_0$
and $\mu_1$, respectively. Since ${\cal K}$ is
lower-semi-continuous in both $\mu_0$ and $\mu_1$ we obtain the
desired result from \er{iiijk} .

\item
 {${\cal
E}(\mu_0,\mu_1)\leq {\cal K}(\mu_0,\mu_1)$}. \\  Let $\lambda\in
C^*(\Omega\times\Omega)$ be an optimizer of ${\cal K}$. Since
$\pi^{(1)}_\#\lambda=\mu_1$ then $$\int_\Omega\phi_1(x)\mu_1(dx)=
\int_\Omega\int_\Omega \phi_1(y)\lambda(dxdy) \ \ \ \text{and} \ \
\int_\Omega\phi_0(x)\mu_0(dx)= \int_\Omega\int_\Omega
\phi_0(x)\lambda(dxdy)$$ for any continuous $\phi_1, \phi_2$. Set
$\phi_1(x)=\psi(x,T)$ and $\phi_0(x)=\psi(x,0)$ with $\psi$ an
optimal  backward solution of problem ${\cal E}$. Then $${\cal E}
= \int_\Omega \psi(x,T)\mu_1(dx)-\int_\Omega\psi(x,0)\mu_0(dx)=
\int_\Omega\int_\Omega\left[ \psi(y,T)-\psi(x,0)\right]
\lambda(dxdy) \ . $$ Since $\psi$ is a backward solution then $$
\int_\Omega\int_\Omega\left[ \psi(y,T)-\psi(x,0)\right]
\lambda(dxdy)\leq \int_\Omega\int_\Omega
J_P(x,y,0,T)\lambda(dxdy)= {\cal K} \ . $$
\end{itemize}
%\end{proof}
\par
\vskip .3in\noindent We have proved
\be\label{underupper}\int_\Omega
\ops(x,T)\mu_1(dx)-\int_\Omega\ops(x,0) \mu_0(dx)= \int_\Omega
\ups(x,T)\mu_1(dx)-\int_\Omega\ups(x,0) \mu_0(dx)= {\cal
L}(\mu_0,\mu_1) \ . \end{equation}
 We now turn to the proof of parts (i)-(vi)
of the Theorem.
%\end{proof}
\begin{description}
\item{i)}
 Let $\mu^{(0)}$ be  a minimizer of ${\cal L}$.
Given $t\in I_0$, let $\mu_{1/2}:= \mu^{(0)}_{(t)}\in {\cal M}$.
Let us consider $\mu^{(1)}$ to be the restriction of $\mu^{(0)}$
to  $\Omega\times [0,t]$ and $\mu^{(2)}$
 the restriction of $\mu^{(0)}$ to  $\Omega\times[t,T]$.
 Evidently, $\mu^{(1)}$ is a minimizer of $L_P$ on the set of
 orbits $\Lambda_2(\mu_0,\mu_{1/2})$ confined to $[0,t]$ while
 $\mu^{(2)}$ is a minimizer  on  $\Lambda_2(\mu_{1/2}, \mu_1)$  with respect
 to the same set, confined to $[t,T]$.
In particular, \be\label{sum12}L_P(\mu^{(1)}) +
L_P(\mu^{(2)})=L_P(\mu^{(0)})={\cal L}(\mu_0,\mu_1) \ .
\end{equation}
 \par
By what we know so far, \be\label{psiup}\int_\Omega
\ops(x,t)\mu_{1/2}(dx)-\int_\Omega\ops(x,0) \mu_0(dx)\leq
L_P(\mu^{(1)}) \end{equation} \be\label{psidow}\int_\Omega
\ops(x,T)\mu_{1}(dx)-\int_\Omega\ops(x,t) \mu_{1/2}(dx)\leq
L_P(\mu^{(2)}) \ . \end{equation} However, if we sum \er{psiup}
and \er{psidow} and use \er{sum12} and \er{underupper}, we
conclude that there is, in fact, an equality in both \er{psiup}
and \er{psidow}.
 Same argument holds for $\ups$ as well.
Thus $$\int_\Omega \ops(x,t)\mu_{1/2}(dx)-\int_\Omega\ops(x,0)
\mu_0(dx)= L_P(\mu^{(1)})=\int_\Omega
\ups(x,t)\mu_{1/2}(dx)-\int_\Omega\ups(x,0)\mu_0(dx)\ . $$ Since
$\ops(x,0)\equiv \ups(x,0)$, $$\int_\Omega
\left[\ops(x,t)-\ups(x,t)\right]\mu_{1/2}(dx)=0 \ . $$ But,
$\ops\geq \ups$ by Lemma~\ref{revpair}. Hence
$\ops(x,t)=\ups(x,t)$ on
$supp\left(\mu^{(0)}_{(t)}\right)=supp\left(\mu_{1/2}\right)$.
This, together with Lemma~\ref{cc1}, proves that
$Supp\left(\mu^{(0)}\right)\cap \Omega_{I_0}\subset K_0\equiv \{
(x,t) \ ; \ t\in I_0 \ ,  \ups(x,t)=\ops(x,t)\}$ and, in
particular, that $\phi$ is differentiable at {\it any} point on
the support of $\mu^{(0)}$ in $\Omega_I$.
\item{ii)}
This part follows from Lemma~\ref{ccc1}. In addition, the limits
$\lim_{\tau\rightarrow T} {\bf T}_t^\tau$ and
$\lim_{\tau\rightarrow 0} {\bf T}_\tau^t$ exists since
$\nabla_x\psi$ is uniformly bounded on $K_0$. In particular, the
Lipschitz extension $\vec{v}$ can be chosen to be a uniformly
bounded function on $\Omega_I$ as well.
\item{iii)}
 Suppose there are two optimal solutions $\psi_1, \psi_2$ of ${\cal E}(\mu_0,\mu_1)$.
 To prove the uniqueness for the vector field $\vec{v}$ we
 claim that
 $$ \int_{\Omega_I}\left|\nabla_x\psi_1-\nabla_x\psi_2\right|^2
 \mu(dxdt)=0$$
 for any minimizer $\mu\in\Lambda_2(\mu_0,\mu_1)$ of ${\cal L}(\mu_0,\mu_1)$.
 Let $\psi=\alpha\psi_1+(1-\alpha)\psi_2$ where $\alpha\in (0,1)$.
 Then
 $$ |\nabla_x\psi|^2 =
 \alpha|\nabla_x\psi_1|^2 + (1-\alpha)|\nabla_x\psi_2|^2 -
 \alpha(1-\alpha) |\nabla_x\psi_1-\nabla_x\psi_2|^2\ , $$
 so $\psi_t+|\nabla_x\psi|^2/2 < P$ and \be\label{qret}\int_{\Omega_I}\left[\psi_t +
 |\nabla_x\psi|^2/2\right]\mu(dxdt)<
  \int_{\Omega_I}P\mu(dxdt)\end{equation} if $\nabla_x\psi_1\not=\nabla_x\psi_2$ at
 {\it some} point in the support of a minimizer $\mu$ (recall that both $\nabla\psi_i$, $i=1,2$ are continuous on the support of $\psi$ by
 Lemma~\ref{cc1}).

 On the other hand,
 \be \label{qret0} {\cal E}(\mu_0,\mu_1) = \int_\Omega \left[\psi(x,T)\mu_1(dx)-
 \psi(x,0)\mu_0(dx)\right]={\cal L}(\mu_0,\mu_1)\end{equation}
follows from the assumptions that both $\psi_1,\psi_2$ are
maximizers of ${\cal E}$. From \er{phi}, \er{qret} and \er{qret0}
it follows that
$$L_P(\mu):= \frac{1}{2}||\mu||^2_2+
\int_{\Omega_I}P\mu(dxdt) >  \frac{1}{2}||\mu||^2_2+
\int_{\Omega_I}(\psi_t+|\nabla_x\psi|^2/2)\mu(dxdt) \geq {\cal
E}(\mu_0,\mu_1)={\cal L}(\mu_0,\mu_1) \ , $$ in contradiction to
the assumption that $\mu$ is a minimizer of $L_P$.
\item{iv) } \
 Let, again, $\mu\in\Lambda_2(\mu_0,\mu_1)$
a minimizer of ${\cal L}$ and $\psi$ a maximizer of ${\cal E}$.
Since $\psi$ satisfies the HJ equation on a closed set $K_0$
containing the support of $\mu$ in $\Omega_{I_0}$ and is a $C^{1}$
function  there, we can extend it as a $C^{1}$ function on
$\Omega_{I_0}$ so $\psi\in C^{1}(\Omega_{I_0})\cap LIP(\Omega_I)$
and, by (\ref{qret0}), \be -\int_{\Omega_I}\left[
\psi_t+\frac{1}{2}|\nabla_x\psi|^2-P\right]\mu(dxdt)+\int_\Omega
\left[\psi(x,T)\mu_1(dx)-
 \psi(x,0)\mu_0(dx)\right]={\cal L}(\mu_0,\mu_1)=L_P(\mu)\label{qret2}\end{equation}
 We now use Corollary~\ref{cor1} (\ref{phi})
to observe that $\psi$ is a maximizer of the left of
(\ref{qret2}),  so by taking the variation $\phi=\psi+\eps\eta$
with $\eta\in C^1(\Omega_I)$ we obtain

$$\int_{\Omega_I}(\eta_t+\nabla_x\psi\cdot\nabla_x\eta)\mu(dxdt) +
\int_\Omega \eta(x,0)\mu_0(dx)-\int_\Omega\eta(x,T)\mu_1(dx)\geq
0$$ for {\it any} such $\eta$. Replacing $\eta$ by $-\eta$ we
obtain the equality above. Moreover, by the same argument
following \er{sum12} to \er{psidow} we also obtain that
\be\label{qret3}\int_{t_0}^t\int_{\Omega}(\eta_t+\nabla_x\psi\cdot\nabla_x\eta)\mu_t(dx)
+ \int_\Omega
\eta(x,t_0)\mu_{(t_0)}(dx)-\int_\Omega\eta(x,t)\mu_{(t)}(dx)= 0 \
,
\end{equation} hold for any $0<t_0<t<T$. In particular, $\mu$
solves the weak form of the continuity equation with
$\vec{v}=\nabla_x\psi$.
\par
Now, we know that, by the additional assumption on $P$, that $K_0$
is invariant with respect to the flow ${\bf T}_{t_0}^{t}$ induced
by the {\it Lipschitz} vectorfield $\vec{v}$ extending
$\nabla_x\psi$. We shall now prove that $\mu$ is transported by
this flow. That is, for any choice of $t_0,t\in (0,T)$, we need to
show that $\mu_{(t)}=\gamma_t$ where
$$\gamma_t:=\left[ {\bf T}_{t_0}^{t}\right]_\# \mu_{t_0}  \ \ .$$
Since ${\bf T}$ is the flow generated by $\vec{v}$ and $K_0$ is
invariant with respect to $\vec{v}$ it follows that
$\gamma=\gamma_tdt$  is supported on $K$ and solves the weak form
of the continuity equation as well. Setting
$\zeta_t=\mu_{(t)}-\gamma_t$, $\zeta:=\zeta_tdt$  we obtain from
(\ref{qret3})
\begin{equation}\label{weakf}\int_{t_0}^t\int_{\Omega}
(\eta_\tau+\vec{v}\cdot\nabla_x\eta)\zeta_\tau(dx)d\tau=
\int_{\Omega} \eta(x,t)\zeta_t(dx) \end{equation} for any $\eta\in
C^1 ([t_0,t]; \R)$  where we used $\zeta_{t_0}\equiv 0$.
\par
Let now $h=h(x)\in C^1(\Omega)$. Let $\eta=\eta(x,\tau)$ be a
solution of
\begin{equation}\label{direct}  \eta_\tau+\vec{v}\cdot\nabla_x \eta=0 \ \ ; \ \
\eta(x,t)=h(x) \ , \ t_0\leq \tau\leq t \ .  \end{equation} Since,
by Lemma~\ref{cc1} and Lemma~\ref{ccc1}, the vector field
$\nabla_x\psi=\vec{v}$ is locally Lipschitz continuous on $K_0$
which is invariant with respect to the induced flow, we can find a
solution of (\ref{direct}) on $K\cap (\Omega\times [t_0,t])$ via
\begin{equation}\label{charact} \eta(x,\tau)=
h\left(T_\tau^t(x)\right) \ .
\end{equation}
The  function $\eta$ so defined can be extended into a  $C^1$
function on $\Omega\times [t_0,t]$. It satisfies \er{direct} on
$K_0$, so, recalling that $\zeta$ is supported on $K_0$, we
substitute now (\ref{direct}) in (\ref{weakf}) to obtain
$\zeta_t\equiv 0$ and the proof of part (iv).
\item{v)} The optimality of ${\bf T}$ is evident from the proof of
(iii) and (iv).
\item{vi)} \ From the last part of Lemma~\ref{createrev} it
follows that $\psi$ is a reversible solution so Lemma~\ref{cc1}
implies that $\psi_t + |\nabla_x\psi|^2/2=0$ is satisfied {\it
everywhere} on $\Omega_{I_0}$. The flow induced by such a solution
is given by ${\bf T}_\tau^t(x)=x+(t-\tau)\nabla_x\psi(x,\tau)$
and, by (iv) and (v), it transports $\mu_{(\tau)}$ to $\mu_{(t)}$
optimally.
\end{description}

\section{Appendix}
\begin{proof} of Proposition~\ref{propdual}: \\  Define $$\Phi(c^*,z):= {\cal F}(c^*) - <c^*,z> + <h,z> \
.
$$ First, note that $$I= \inf_{c^*\in {\bf C}^*} \sup_{z\in {\bf
Z}} \Phi(c^*,z) \ . $$ Indeed, if $c^*\not\in {\bf Z}^*$ then
$\sup_{z\in {\bf Z}}\Phi(c^*,z)=\infty$ while, if $c^*\in {\bf
Z}^*$ then $\phi(c^*,z)={\cal F}(c^*)$ by definition. We have,
therefore, to show $$ \inf_{c^*\in {\bf C}^*} \sup_{z\in {\bf Z}}
\Phi(c^*,z)=  \sup_{z\in {\bf Z}} \inf_{c^*\in {\bf
C}^*}\Phi(c^*,z) \ .  $$ It is trivial that $$ \inf_{c^*\in {\bf
C}^*} \sup_{z\in {\bf Z}} \Phi(c^*,z)\geq   \sup_{z\in {\bf Z}}
\inf_{c^*\in {\bf C}^*}\Phi(c^*,z):=\underline{I} \ ,  $$ so we
only have to show that \be\label{appfin} \inf_{c^*\in {\bf C}^*}
\sup_{z\in {\bf Z}} \Phi(c^*,z)\leq \underline{I} \ .
\end{equation} Define, for any $z\in {\bf Z}$ $$ A_z=\left\{
c^*\in {\bf C}^* \ ; \ \Phi(c^*,z)\leq \underline{I}\right\} \ .
$$
 Note that (\ref{appfin}) follows provided
 \be \label{appfin1}\bigcap_{z\in {\bf Z}} A_z\not=\emptyset  \ . \end{equation}
 The next step is to show that, for any finite set $z_1, \ldots
 z_n\in {\bf Z}$, the set $\bigcap_{z_i} A_{z_i}\not=\emptyset$.
 The proof of this part  can be taken from the proof of Theorem 2.8.1 in
 [Ba].
 \par
Finally, note that $A_0\subset \overline{A}_0$ as defined in the
Proposition, since $\underline{I}\leq I$. It follows that $A_0$ is
compact, and that the non-empty intersection of finite sets
implies (\ref{appfin1}). \end{proof}
\vskip .3in \noindent
\begin{proof} of Lemma~\ref{lemma3.25}: \\
 Let $\rho_1(r)$ be a smooth, positive function with compact
support such that $$| \mathbb{S}^{n-1}|\int_0^\infty r^{n-1}
\rho_1(r)dr = 1 \ \ ; \ \ \int_0^\infty r^{k}\rho_1(r) dr:= M_k \
\ ; \ \  \int_0^\infty r^{n-1}\rho_1^p dr := L(p) \ . $$ Set also
$$\rho_\alpha(r)=\alpha^n\rho_1(\alpha r) \ \ . $$
 Define
 $$ {\bf v}(x,t)=\left\{
 \begin{array}{c}
   \frac{x-\overline{x}(t)}{t-t_0} + \dot{\overline{x}}(t) \ \ \  \text{if} \  t_0\leq t\leq (t_0+t_1)/2 \\
   \frac{x-\overline{x}(t)}{t_1-t} + \dot{\overline{x}}(t)  \ \ \ \  \text{if} \  (t_0+t_1)/2\leq t\leq
   t_1
 \end{array}\right.$$

 $$\rho(x,t)=\left\{ \begin{array}{c}
    \frac{1}{(t-t_0)^n}\rho_\alpha\left(
 \frac{|x-\overline{x}(t)|}{t-t_0}\right) \ \ \ \text{if} \ \  \ t_0\leq t\leq (t_0+t_1)/2 \\
    \frac{1}{(t_1-t)^n}\rho_\alpha\left(
 \frac{|x-\overline{x}(t)|}{t_1-t}\right) \ \ \ \text{if} \ \  \ (t_0+t_1)/2\leq t\leq t_1 \
 \end{array}\right.
  \ . $$
 A
direct calculation shows that $\rho$ satisfies the weak form of
the continuity equation: $$\rho_t + \nabla_x\cdot \left({\bf
v}\rho\right)=0 \ .
$$
Let us now consider the interval $[t_0, (t_0+t_1)/2]$. The second
interval $[(t_0+t_1)/2, t_1]$ can be treated analogously.
  Define the lifting of $\rho$ as $$f(x,t,v)=
\sigma^{-n}\pi^{-2/n} \exp \left(-\frac{|v-{\bf
v}|^2}{\sigma^2}\right)\rho(x,t) \ . $$ It follows immediately
that $$\int_{\R^n} v^2 f(x,t,v)dv= \frac{\sigma
n}{2}\rho(x,t)+{\bf v}^2\rho(x,t) \ \ ; \ \ \int_{\R^n} |f|^p
dv=p^{-n/2}\pi^{2(1-p)/n} \sigma^{n(1-p)}\rho^p(x,t) \ . $$
Moreover: $$ \int_\Omega \rho(x,t) dx=1 \ \ ; \ \ \int_\Omega
\rho^p(x,t) dx= (t-t_0)^{n(1-p)}| \mathbb{S}^{n-1}|\int_0^\infty
r^{n-1}\rho^p_\alpha(r) dr=\alpha^{n(p-1)}(t-t_0)^{n(1-p)}|
\mathbb{S}^{n-1}|L(p)$$ $$ \int_{\Omega} |{\bf v}(x,t)|^2\rho(x,t)
dx=\left|\dot{\overline{x}}(t)\right|^2  +|\mathbb{S}^{n-1}|
\int_0^\infty r^{n+1} \rho_\alpha(r)dr =
\left|\dot{\overline{x}}(t)\right|^2 +|\mathbb{S}^{n-1}|
\alpha^{-2}M_{n+1} $$

In particular: $$\int_\Omega\int_{t_0}^{(t_0+t_1)/2}|v|^2f=
\int_{t_0}^{(t_0+t_1)/2}\left|\dot{\overline{x}}\right|^2 dt +C_1
|t_1-t_0|\alpha^{-2} +
 O(\sigma) $$ and
 $$\int_\Omega\int_{t_0}^{(t_0+t_1)/2}\rho^p=
C_3 |t_1-t_0|^{n(1-p)+1} \alpha^{n(p-1)} $$
\end{proof}
\vskip .3in\noindent
\begin{proof} {\it of Lemma~\ref{holderd}} \\
We use Corollary~\ref{cor1} with $\mu$ supported on
$\Omega\times[t_1, t_0]$ and $\mu_{t_0}=\delta_{x_0}$,
$\mu_{t_1}=\delta_{x_1}$, to obtain
\be\label{phix1x0}\phi(x_1,t_1)-\phi(x_0,t_0) \leq
\frac{1}{2}||\mu||^2_2 + \left| \int_{\Omega}\int_{t_0}^{t_1}
(\phi_t+|\nabla_x\phi|^2/2) \mu_{(t}(dx) dt\right| \ .
\end{equation} By Lemma~\ref{lemma3.25} we can find such a $\mu$
for which: \be \label{meq}||\mu||^2_2\leq
\int_{t_0}^{t_1}\left|\dot{\overline{x}}\right|^2 +C_1 |t_1-t_0|
\alpha^{-2} \end{equation} and, for the density $\rho=\rho_\mu$:
 \be\label{rhoeq}\int_\Omega\int_{t_0}^{t_1}\rho^p\leq
C_3 |t_1-t_0|^{n(1-p)+1} \alpha^{n(p-1)} \end{equation} where $p <
1+1/n$ and $\alpha$  any positive constant.  Since $\rho$ is
supported, for any $t$, in a domain of diameter
$(t_1-t_0)\alpha^{-1}$ it follows $$\int_{\Omega}\int_{t_0}^{t_1}
(\phi_t+|\nabla_x\phi|^2/2) \rho dx dt = O\left(\frac{L
(t_1-t_0)^2}{\alpha}\right) + \int_{t_0}^{t^1}
P(\overline{x}(t),t)dt +\int_{\Omega}\int_{t_0}^{t_1} \xi \rho dx
dt\ , $$ where $L$ is the Lipschitz norm of $P$.  By (\ref{rhoeq})
we obtain \be\label{last} \left|\int_{\Omega}\int_{t_0}^{t_1} \xi
\rho dx dt\right|
 \leq ||\rho||_p
||\xi||_s \leq C_3^{1/p}|t_1-t_0|^{[n(1-p)+1]/p }\alpha^{n(p-1)/p}
||\xi||_s \ . \end{equation} Collecting (\ref{phix1x0}) to
(\ref{last})
 $$
\phi(x_1,t_1)-\phi(x_0,t_0) \leq
\frac{1}{2}\int_{t_0}^{t_1}\left|\dot{\overline{x}}\right|^2 dt
 + C||\xi||_s
  |t_1-t_0|^{[n(1-p)+1]/p}\alpha^{n(p-1)/p}
 +C_1 |t_1-t_0| \alpha^{-2} +O\left(\frac{L (t_1-t_0)^2}{\alpha}\right) \ . $$
 The choice $\alpha=||\xi||_s^{-\beta}(t_1-t_0)^{\gamma}$ where $\gamma =
\frac{(n+1)(p-1)}{2p+n(p-1)}$  and $\beta=\frac{p}{2p+n(p-1)}$ is
the optimal choice and yields the desired result.
\end{proof}
\newpage

\vskip .3in
\begin{center}{\bf References}\end{center}
\begin{description}
\item{[Am]} \ L. Ambrosio: {\it Lectures Notes on Optimal Transport
Problems}, CVGMT Preprint: \\ http://cvgmt.sns.it/papers/amb00a/
\item{[AGS]} \ \ L. Ambrosio, N. Gigli \& G. Savare: {\it
Gradient flows of probability measures}, Preprint.
\item{[AFP]} \  \ L. Ambrosio, N.Fusco \& D.Pallara: {\it
Functions of Bounded Variations and Free Discontinuity Problems},
Oxford University Press, 2000.
\item{[B]} \ Y. Brenier: {\it Polar factorization and monotone
rearrangement of vector valued  functions}, Comm. Pure Appl. Math,
{\bf 44}, (1991), 375-417.
 \item{[Ba]} \ A.V. Balakrishnan, {\it
Applied Functional Analysis}, Applications of Mathematics 3,
Springer-Verlag, 1976.
\item{[BB]} \ J.D.Benamou, Y. Brenier: {\it A computational fluid
mechanics solution to the Monge-Kantorovich mass transfer
problem}, Numer.Math., {\bf 84} (2000), 375-393.
\item{[BBG]} \ J.D.Benamou, Y. Brenier \& K.Guitter: {\it The Monge-Kantorovich mass transfer
and its  computational fluid mechanics formulation}, Inter. J.
Numer.Meth.Fluids, {\bf 40} (2002), 21-30.
\item{[C]} \ L. Caffarelli, {\it Allocation maps with general cost
functions}, in  Partial Differential Equations with Applications
(ed.  by Talenti), (1996), Dekker
\item{[E]} \ L.C.Evans, {\it  Partial Differential Equations}
 1949-
Providence, R.I. : American Mathematical Society, c1998.
\item{[GM]} \ W. Gangbo \& R.J. McCann: \ {\it The geometry of
optimal transportation}, Acta Math., 177 (1996), 113-161
\item{[M]} \ G. Monge: {\it M\'{e}moire sur la th\'{e}orie des d\'{e}blais et de remblais}, Histoire de
 l'Acad\'{e}mie Royale
des Sciences de Paris, 1781, pp. 666-704
\item{[RKF]} J. Richard, D.  Kinderlehrer \& F. Otto: {\it The variational formulation of the Fokker-Planck
   equation}. SIAM J. Math. Anal. {\bf 29} (1998), no. 1, 1--17
\item{[V]} C. Villani: {\it Topics in Optimal Transportation},
Graduate studies in Math, {\bf 58}, AMS, 2003
\item{[W]} G. Wolansky: {\it Rotation numbers for measure-valued circle
maps}, J. D'anal.  Math., to appear.
\end{description}

\end{document}